\documentclass[aps,prb,floatfix,superscriptaddress,twocolumn]{revtex4-2}
\usepackage{graphics,bm,amsmath,amssymb,natbib,url,epsfig,psfrag, color}
\usepackage{braket}
\usepackage{xcolor}
\usepackage{physics}
 
\usepackage[normalem]{ulem}
\newcommand{\Renyi}[0]{R\'{e}nyi~}
\newcommand{\be}{\begin{equation}}
\newcommand{\ee}{\end{equation}}
\newcommand{\bea}{\begin{eqnarray}}
\newcommand{\eea}{\end{eqnarray}}
\newcommand{\intz}{\mathbb{Z}}

\usepackage{hyperref}
\usepackage{cancel}

\begin{document}
	
	\title{Symmetry-resolved entanglement of 2D symmetry-protected topological states}
	\author{Daniel Azses}
	\affiliation{School of Physics and Astronomy, Tel Aviv University, Tel Aviv 6997801, Israel}
	
	\author{David F. Mross}
	\affiliation{Department of Condensed Matter Physics, Weizmann Institute of Science, Rehovot 7610001, Israel}
	
	\author{Eran Sela}
	\affiliation{School of Physics and Astronomy, Tel Aviv University, Tel Aviv 6997801, Israel}
	
	\begin{abstract}
	Symmetry-resolved entanglement is a useful tool for characterizing symmetry-protected topological states. In two dimensions, their entanglement spectra are described by conformal field theories but the symmetry resolution is largely unexplored. However, addressing this problem numerically requires system sizes beyond the reach of exact diagonalization. Here, we develop tensor network methods that can access much larger systems and determine universal and nonuniversal features in their entanglement. Specifically, we construct one-dimensional matrix product operators that encapsulate all the entanglement data of two-dimensional symmetry-protected topological states. We first demonstrate our approach for the Levin-Gu model. Next, we use the cohomology formalism to deform the phase away from the fine-tuned point and track the evolution of its entanglement features and their symmetry resolution. The entanglement spectra are always described by the same conformal field theory. However, the levels undergo a spectral flow in accordance with an insertion of a many-body Aharonov-Bohm flux.

	\end{abstract}
	
	\maketitle
	\section{Introduction} 
		Symmetry-protected topological states (SPTs) are characterized by a symmetric bulk state that does not host fractional excitations. Still, they are topological in the sense of carrying anomalous edge states at their boundary with a trivial state or different SPTs. In two dimensions, the edges are described by a one-dimensional conformal field theory (CFT) \cite{chen2012chiral,levin2012braiding,santos2014symmetryprotected,scaffidi2016wave,han2017boundary}. The presence of these states is dictated by a specific structure in their entanglement. Yet, unlike topologically-ordered (fractional) states, the entanglement entropy of SPTs does not contain a topological term. Instead, the topological nature of these states may be revealed by resolving the entanglement entropy according to symmetries or by studying entanglement spectra (ES).

The entanglement entropy of a system with global symmetries can be decomposed according to the associated quantum numbers \cite{goldstein2018symmetry,xavier2018equipartition,bonsignori2019symmetry,calabrese2021symmetry}. Specifically, it is given by the sum of entropies for each choice of these quantum numbers \textit{in one subsystem}.
The \Renyi moments of the symmetry-resolved entanglement are experimentally measurable~\cite{islam2015measuring,neven2021symmetry,vitale2022symmetry,rath2022entanglement} as also demonstrated for one-dimensional SPT states~\cite{azses2020identification,azses2021observing} on IBM quantum computers. 	For such states, each symmetry sector contributes equally to the total entropy~\cite{azses2020symmetry}. This equipartition corresponds to exact degeneracies in the ES~\cite{pollmann2010entanglement,cornfeld2019entanglement}. These have been recognized as the source of the computational power of one-dimensional SPTs~\cite{else2012symmetry} within measurement-based quantum computation~\cite{raussendorf2001one}.

The ES generalizes the entanglement entropy and contains additional universal information. For 2D topological states with a chiral edge, the Li-Haldane conjecture~\cite{li2008entanglement,qi2012general} states that the levels of the ES correspond to the conformal field theory that describes a physical edge. Extrapolating to the nonchiral case, one may expect that the ES of SPT phases have the same universal properties as their nonchiral edge CFTs, such as the central charge $c$. For example, for SPTs stabilized by a $\intz_N$ symmetry, the free boson CFT with $c=1$ was found to describe the edge~\cite{levin2012braiding,chen2012chiral}. While the ES should correspond to the same CFT as that of the physical edge, the two may differ by nonuniversal parameters such as the compactification radius. Other nonuniversal properties include effective fluxes, which change the boundary condition of the 1D edge.

Our motivating questions are: how does the ES decompose according to symmetry in 2D SPTs? How does this decomposition fit into the CFT? What are its universal properties, and how does the ES vary within a given SPT phase? Previous work by Scaffidi and Ringel exploring the emergent CFT in the ES of SPTs was limited to small system sizes~\cite{scaffidi2016wave}. It could, in principle, have performed a symmetry resolution, which does not require large systems. By contrast, distinguishing universal from nonuniversal properties upon continuous variation of the ground states, as we find, does require large system sizes.

In order to address these issues in this work, we develop an efficient numerical method~\cite{zou2016spurious,arildsen2022entanglement} to calculate the ES of short-range entangled states in two dimensions. Our method is summarized in Fig.~\ref{fig:schamatics} and described in Sec.~\ref{se:general_construction}. It uses a quantum circuit representation to construct gapped one-dimensional models that exhibit the same entanglement properties as the 2D SPTs in question. In particular, they allow us to extract the entanglement spectra of large systems and their symmetry resolution using tensor network methods~\cite{verstraete2008matrix,eisert2013entanglement,orus2014practical,feldman2022entanglement}.

In Sec.~\ref{se:Levin_Gu}, we apply this approach to the Levin-Gu model~\cite{levin2012braiding} on an infinite cylinder with circumferences as large as $L=150$. In agreement with previous studies of much smaller systems \cite{scaffidi2016wave}, we observe the spectrum of a CFT with central charge $c=1$. Specifically the ES can be organized in terms of primary states and their descendants. We find that all the descendant states have the same subsystem symmetry quantum numbers as the corresponding primary state. We also identify an unexpected subtlety in the ES of the Levin-Gu model: it distinguishes cylinders whose circumference is a multiple of three from all others. We attribute this effect to the lattice and show that it translates into a flux insertion of the corresponding CFT.

In Sec.~\ref{se:MPS_cocycle}, we apply our method to explore more generic states. 
We construct a continuous family of wave-functions within one SPT phase using the framework of cohomological classification~\cite{chen2013symmetry}. The ES of these states reveal a direct relation between so-called coboundary transformations and certain fluxes affecting the many-body SPT states.

In Sec.~\ref{se:H1DLevin_Gu}, we further elaborate on the gapped one-dimensional models derived in Sections \ref{se:Levin_Gu} and \ref{se:MPS_cocycle}. We demonstrate that they can be used to obtain the central charge of the SPT edge very efficiently without reference to ES. Restricting the edge Hamiltonian to terms that act within a single subsystem results in a critical chain that is described by the same CFT as the SPT edge. The central charge of such one-dimensional chains can be readily
extracted from the scaling of the entanglement entropy, which satisfies the Calabrese-Cardy formula~\cite{clabrese2004entanglement}.

		\section{Dimensional reduction and tensor network approach}
		\label{se:general_construction} 

		\begin{figure}[t]
			\centering
			\includegraphics[width=\linewidth]{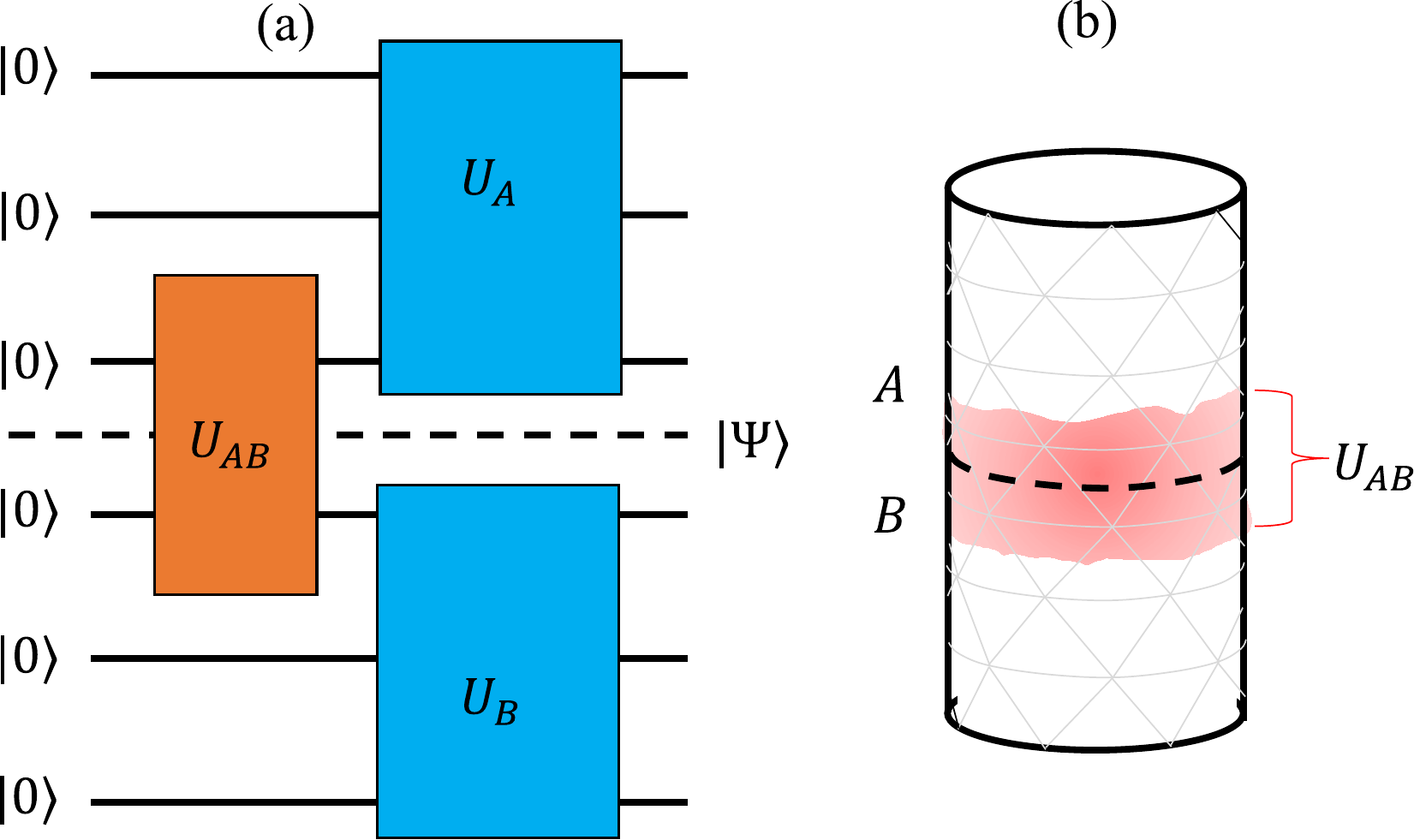}
			\caption{Schematics of our method: (a) We analyze a state $| \Psi \rangle$ given as a quantum circuit representation as in Eq.~(\ref{eq:UAUBUAB}). The entanglement between subsystems $A$ and $B$ is created by a unitary $U_{AB}$ acting within a finite distance from the entanglement cut (dashed line). (b) We consider an SPT on a cylinder consisting of regions $A$ and $B$.}
			\label{fig:schamatics}
		\end{figure}
	
	Before specifying to 2D, consider a $d$-dimensional space. The two subsystems $A$ and $B$ share a $(d-1)$-dimensional boundary $\partial A=\partial B$. By virtue of their finite depth circuit representation~\cite{chen2011classification}, any SPT state can be written as 
		\be
		\label{eq:UAUBUAB}
		|\Psi \rangle = U_A U_B U_{AB}| 0 \rangle .		
		\ee
	Here $| 0 \rangle$ is a site-factorizable product state, i.e., the ground state of a trivial gapped Hamiltonian $H_0$ that is the sum over one-site operators. The unitaries $U_{A}$ and $U_{B}$ act only on subsystems $A$ and $B$, respectively. $U_{AB}$ acts in a $(d-1)$-dimensional region denoted $C_{AB}$ extending a finite distance from $\partial A$; see Fig.~\ref{fig:schamatics}(a). Consequently, the ES is fully encoded in $U_{AB}$. Indeed, the reduced density matrix of region A is given by
		\bea
		\label{eq:rhoA}
		\rho_A ={\rm{Tr}}_B | \Psi \rangle \langle \Psi | = U_A {\rm{Tr}}_B \left( U_{AB} |0 \rangle \langle 0 | U_{AB}^\dagger \right) U_A^\dagger. 
		\eea
Up to the unitary transformation $U_A$, which does not affect the ES, $\rho_A$ acts nontrivially only in the interface region $C_{AB}$. Consequently, the ES can be fully encoded by a state 
$| \psi_{{\rm edge}} \rangle$ that lives only in $C_{AB}$. We define this edge state via $U_{AB} |0 \rangle = | \psi_{{\rm edge}} \rangle \otimes |0\rangle^{A \bigcup B \smallsetminus C_{AB}} $ such that the operator 
\bea
\rho_{{\rm edge}}={\rm{Tr}}_B |\psi_{{\rm edge}} \rangle \langle \psi_{{\rm edge}} |
\eea 
exhibits the same spectrum as $\rho_A$.
The `edge density matrix' $\rho_{{\rm edge}}$ acts nontrivially only on a $(d-1)$-dimensional region within subsystem $A$. It is the central object in this paper, which we construct using tensor-network methods; see Fig.~\ref{fig:schamatics2}(c). 

The presence of an on-site symmetry generated by $S = S_A \otimes S_B $, implies that $[\rho_A,S_A]=0$.
It follows that $[\rho_{{\rm edge}},U_A^\dagger S_A U_A]=0$. As a result, the symmetry-resolved ES is obtained by diagonalizing $\rho_{{\rm edge}}$ simultaneously with the edge symmetry operator 
\be
\label{se:S}
S_{{\rm{edge}}}=U_A^\dagger S_A U_A.
\ee

It is convenient to think of the pure state $| \psi_{{\rm edge}} \rangle$ as the ground state of a local {\emph{edge Hamiltonian}} 
\be
H_{{\rm edge}} =U_{AB} H_0 U_{AB}^\dagger, 
\label{hedge}
\ee
which has the same (gapped) spectrum as $H_0$.
$H_{{\rm edge}}$ acts on the support of $C_{AB}$, which contains the union of two adjacent $(d-1)$-dimensional regions in $A$ and $B$.
Accordingly, $H_{{\rm{edge}}}$ can be separated as 
\be
\label{eq:HABAN}
H_{{\rm edge}}=H_A+H_B+H_{AB}.
\ee
 We remark that, unlike the standard ``entanglement Hamiltonian" $H_E$ defined by $\rho_A=e^{-H_E}$, the edge Hamiltonian $H_{\rm{edge}}$ acts on \textit{both} subsystems. However, in Sec.~\ref{se:H1DLevin_Gu}, we argue that $H_E$ and $H_A$ describe a CFT with the same universal properties dictated by the bulk SPT phase.

\section{Symmetry-resolved ES of the Levin-Gu Model}
\label{se:Levin_Gu} 
Next, we focus on the paradigmatic 2D Levin-Gu model and demonstrate our method for the computation of the ES and its symmetry resolution.

The Levin-Gu state admits a quantum circuit form~\cite{liu2020many}
\be
\label{eq:LG}
|\Psi_\text{LG} \rangle = U_{CCZ} U_{CZ}U_{Z} |+ \rangle,
\ee
where $U_{CCZ}$, $U_{CZ}$, and $U_Z$ are, respectively, the products of $CCZ$, $CZ$ and $Z$ gates acting on all triangles, edges, and vertices, and $|+ \rangle$ is the ground state of $H_0=-\sum_i X_i$. We consider a cylinder geometry as displayed in Fig.~\ref{fig:schamatics}(b). Next, we use this quantum circuit form to derive a 1D Hamiltonian that encodes the ES of this 2D model, exemplifying the general prescription of Sec.~\ref{se:general_construction}.

\subsection{Gapped 1D edge Hamiltonian}

		\begin{figure}[t]
			\centering
			\includegraphics[width=\linewidth]{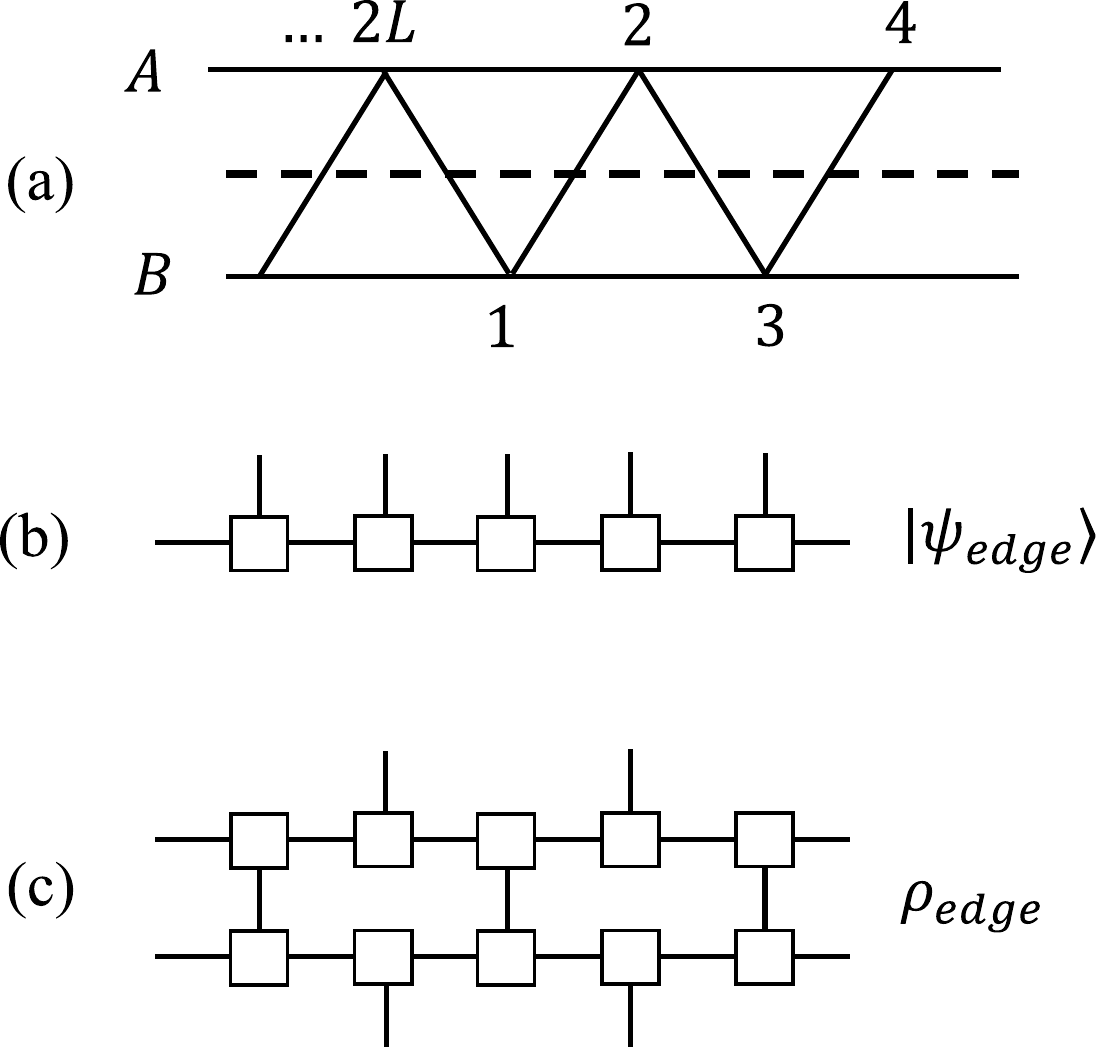}
			\caption{(a) The resulting 1D edge Hamiltonian corresponding to Fig.~\ref{fig:schamatics}(b) with the underlying triangular lattice is a zigzag chain containing both $A$ (even) and $B$ (odd) sites. (b) We construct a MPS of the ground state of $H_{{\rm{edge}}}$ living on $A$ and $B$. (c) The reduced density matrix of subsystem $A$ is constructed as an MPO by contracting the odd sites $(\in B)$ of two copies of the MPS state. }
			\label{fig:schamatics2}
		\end{figure}

The Levin-Gu state provides an explicit example of Eq.~(\ref{eq:UAUBUAB}). In this case,
\be
\label{eq:UA}
U_A=\prod_{\text{triangles}\in A} U_{CCZ}^{ijk} \prod_{\text{links} \in A} U^{ij}_{CZ} \prod_{\text{sites} \in A} U_{Z}^{i},
\ee
and similarly for $U_B$. The triangles and links 
that connect the two subsystems identify the interface region $C_{AB}$ as a zigzag chain with $i~{\rm{even}} \in A$ and $i~{\rm{odd}} \in B$; see Fig.~\ref{fig:schamatics2}(a).
The corresponding entangling gates are
\be
U_{AB} = \prod_{i} U_{CCZ}^{i-1,i,i+1} \prod_{i} U_{CZ}^{i,i+1} \equiv U_{AB}^{CCZ} U_{AB}^{CZ}.
\ee
The edge Hamiltonian of Eq.~\eqref{hedge} is then given by
\be
\label{eq:cluster}
H_{{\rm edge}}=U_{AB} H_0 U_{AB}^\dagger=U_{AB}^{CCZ} H_{{\rm{cluster}}} U_{AB}^{CCZ},
\ee
where $H_{{\rm{cluster}}} = U_{AB}^{CZ} H_0 U_{AB}^{CZ}=-\sum_{i} Z_{i-1} X_i Z_{i+1}$ is the 1D cluster Hamiltonian. A more explicit form of $H_\text{edge}$ is readily obtained by virtue of the identity
\be
U_{CCZ}^{ijk}X_i U_{CCZ}^{ijk}=\frac{1}{2}X_i [I+Z_j+Z_k-Z_j Z_k].
\ee
We thus obtain $H_{{\rm edge}}$ as a sum of tensor products of 1-qubit operators, 
\bea
\label{eq:H1Dedge}
H_{{\rm edge}}=\sum_i \frac{1}{4} X_i [I-Z_{i-2}+Z_{i-1}+Z_{i+1}-Z_{i+2} \nonumber \\
+Z_{i-2}Z_{i-1}+Z_{i-1}Z_{i+1}+Z_{i+1}Z_{i+2} \nonumber \\-Z_{i-2}Z_{i+1}-Z_{i-1}Z_{i+2}-Z_{i-2}Z_{i+2} \nonumber \\
+Z_{i-2}Z_{i-1}Z_{i+1}+Z_{i-2}Z_{i+1}Z_{i+2} \nonumber \\+Z_{i-2}Z_{i-1}Z_{i+2}+Z_{i-1}Z_{i+1}Z_{i+2} \nonumber \\
-Z_{i-2}Z_{i-1}Z_{i+1}Z_{i+2}].
\eea
We construct a matrix product state (MPS) of the ground state of $H_{{\rm edge}}$, denoted $|\psi_{{\rm edge}} \rangle$ and depicted in Fig.~\ref{fig:schamatics2}(b), using the iTensor and Julia libraries~\cite{fishman2020itensor}. The ground state $|\psi_{{\rm edge}} \rangle$ converges with bond dimension $\chi=9$ for periodic boundary conditions.
Subsequently, we construct the matrix product operator (MPO) for $\rho_{{\rm edge}}$ by contracting the $B$ sites of the outer product $|\psi_{{\rm edge}} \rangle \langle \psi_{{\rm edge}} |$; see Fig.~\ref{fig:schamatics2}(c). Finally, an excited state density matrix renormalization group (DMRG) calculation on $\rho_{{\rm edge}}$ yields the ES.

\subsection{Symmetry resolution}
The $\intz_2$ symmetry operator of the Levin-Gu model is $X=\prod_i X_i$. To see that the Levin-Gu state is an eigenstate of $X$ and determine its eigenvalue, we use the quantum circuit form [cf.~Eq.~\eqref{eq:LG}] along with the identities
\bea
\label{eq:identies}
X Z_i X&=&-Z_i, \\
X U_{CZ}^{i j} X&=&-U_{CZ}^{i j} Z_i Z_j, \nonumber \\
X U_{CCZ}^{i j k} X&=&-U_{CCZ}^{i j k} U_{CZ}^{i j} U_{CZ}^{jk}U_{CZ}^{k i} Z_i Z_j Z_k. \nonumber
\eea
The third identity implies, in particular, that
\begin{align}
X U_{CCZ} X =(-1)^T U_{CCZ}~,
\end{align}
where $T$ is the number of triangles. The product over all triangles includes each link and each site an even number of times such that all $Z$ and $CZ$ factors cancel. Similarly, the product over all links includes each site an even number of times, and thus 
\begin{align}
X U_{CZ} X =(-1)^L U_{CZ},
\end{align}
where $L$ is the total number of links. It follows that
\be
X |\Psi_\text{LG} \rangle =(-1)^{T+L+V}|\Psi_\text{LG} \rangle,
\ee
where $V$ is the total number of vertices (i.e., sites). For a perfect triangular lattice without a boundary, $(-1)^{T+L+V}=1$.

According to Eq.~(\ref{se:S}), the edge symmetry operator is given by $S_{{{\rm{edge}}}}^\text{LG} =U_A^\dagger X_A U_A$, where $X_{A}=\prod_{i \in A}X_i$ and $U_A$ is given by Eq.~(\ref{eq:UA}). Unlike the case of the full system, the product over all triangles in subsystem $A$ involves the links along the edge only once. Consequently, commuting $X_A$ across $U_A$ using Eq.~(\ref{eq:identies}) produces uncanceled $Z$ and $CZ$ factors. Consequently, $S_{{{\rm{edge}}}}^{LG}$ acts nontrivially near the edge and we write $U_A^\dagger X_A U_A = S_{{{\rm{edge}}}}^{LG} \otimes \prod_{i \in A \smallsetminus \partial A}X_i$. The first factor only acts on $\partial A $, which contains all even sites. It is given by
\be
S_{{{\rm{edge}}}}^\text{LG} =\prod_{i = {\rm{even}}} X_i \prod_{i = {\rm{even}}} U_{CZ}^{i,i+2} Z_i,
\ee
up to an additional overall factor $(-1)^{T_A+L_A+V_A}=1$ accounting for the total number of triangles, links, and vertices in subsystem $A$. As we can see, in addition to the on-site factor $\prod_{i = {\rm{even}}} X_i$, there is a non-on-site factor~\cite{chen2012chiral}. The latter is the manifestation of the nontrivial SPT. In fact, it is this factor, which is classified~\cite{chen2012chiral} by the third cohomology group $\mathcal{H}^3(\intz_2,U(1))=\intz_2$. One can rewrite the edge symmetry operator as
\be
\label{eq:Sedge}
S_{{{\rm{edge}}}}^{LG} =\prod_{j = {\rm{even}}} X_j \prod_{j = {\rm{even}}} e^{i \frac{\pi}{4} (Z_j Z_{j+2}-1)},
\ee
which shows explicitly that the on-site and non-on-site factors commute.

Having constructed the edge symmetry Eq.~(\ref{eq:Sedge}), we confirm that the eigenstates $|\psi_i \rangle$ of $\rho_{{\rm{edge}}}$ that we obtained from DMRG satisfy $S_{{{\rm{edge}}}}^{LG}|\psi_i \rangle=s_i |\psi_i \rangle$, where $s_i=\pm 1$ is the symmetry eigenvalue.

\subsubsection{Entanglement Hamiltonian}

The above algorithm yields the eigenvalues $\lambda_i$ of the reduced density matrix $\rho_{{\rm edge}}
$, from which we obtain a list of quasienergies $\xi_i =- \log \lambda_i$, being the eigenvalues of the entanglement Hamiltonian defined by $\rho_{{\rm edge}}=e^{-H_E}$. According to the Li-Haldane conjecture~\cite{li2008entanglement}, in topological systems, the latter displays the spectrum of a physical edge. In the present SPT, this spectrum is known to be a nonchiral free boson CFT~\cite{chen2012chiral}.

We match the list of quasienergies $\{\xi_0,\xi_1,\dots \}$ in increasing order to the form
\be
\label{eq:scalingandshifting}
\xi_i-\xi_0=\frac{v}{L} \Delta_i,
\ee
where $v$ is a free parameter corresponding to the velocity of the CFT, $L$ is the circumference of the cylinder, and $\Delta_i$ are ``scaling dimensions" of the CFT, as given below.

\subsection{Entanglement spectrum for $L$ divisible by 3}

As discussed in detail in Appendix~\ref{app:numericalLevinGu}, we can see that the numerically obtained eigenvalues of the entanglement Hamiltonian approximate the free boson spectrum
\be
\label{eq:R}
\Delta(\ell,m,R)=\frac{\ell^2}{R^2}+\frac{R^2 m^2}{4} +{\rm{integers}}\quad (\ell,m \in \mathbb{Z}),
\ee
with compactification radius $R=\sqrt{2}$. As reviewed in Appendix~\ref{se:fieldtheory}, this spectrum can be viewed as an infinite set of primary states $|\ell,m\rangle$. Each of these generates an infinite tower of descendant states denoted in Eq.~(\ref{eq:R}) by ``integers".

The low-lying levels of this spectrum can already be seen to match this pattern in a short system of $L=12$, as seen in Fig.~\ref{fig:es_levingu_12_pbc}. Results for longer systems, shown in Appendix~\ref{app:numericalLevinGu}, confirm this pattern for higher energy levels.
\begin{figure}[t]
	\centering
	\includegraphics[width=\linewidth]{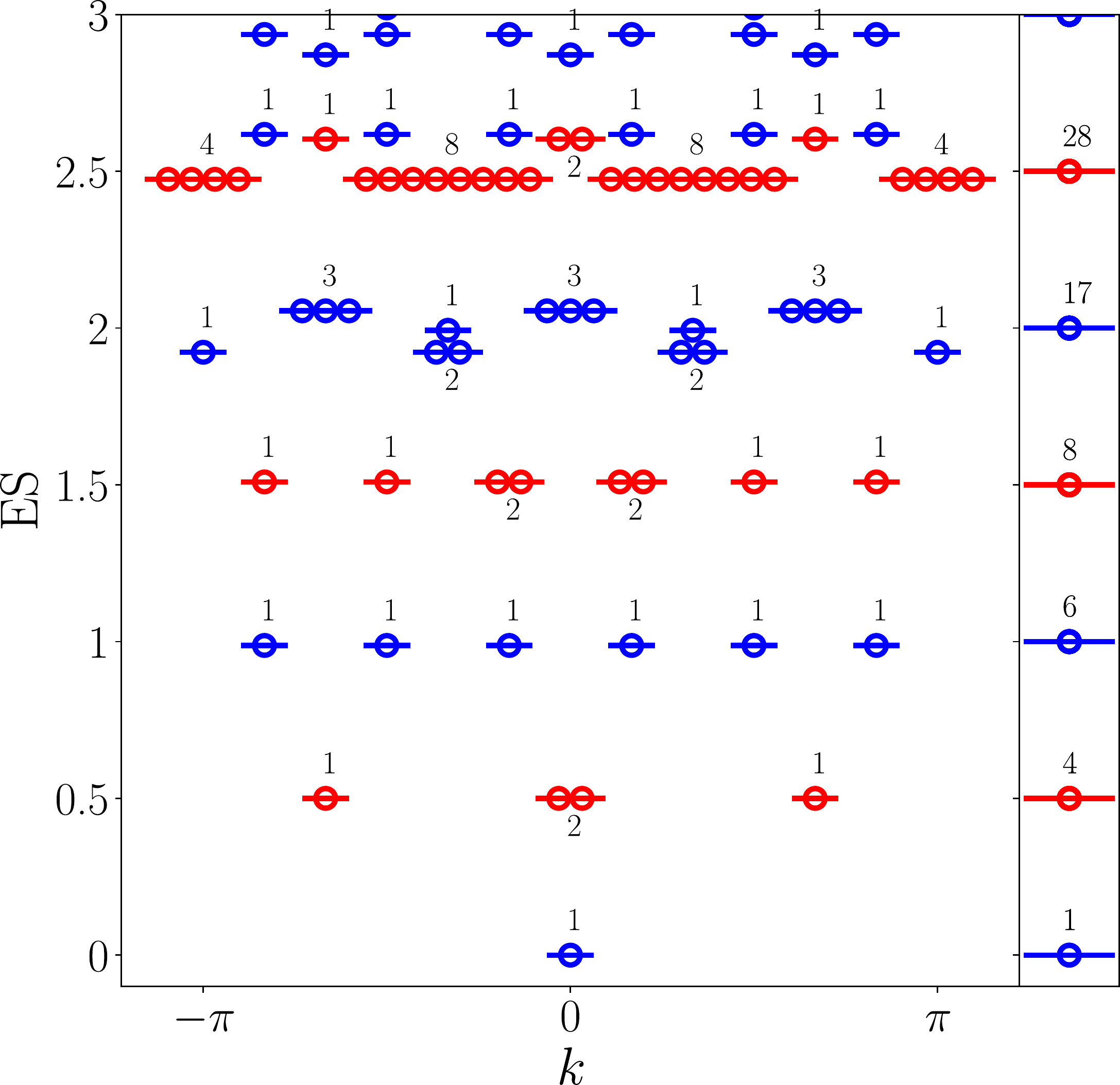}
	\caption{Entanglement spectrum of the Levin-Gu model on a cylinder with circumference $L=12$ and its symmetry resolution. The low-lying energy levels and their degeneracies match Eq.~\eqref{eq:R}, with the second term ``$\rm{integers}$" being given by $\sum_{i>0} i(n_i+\bar{n}_i)$ with $n_i,\bar{n}_i \in \mathbb{N}$ being the $i$'th integer ($i\geq 1$) excitation on top of the primary field denoted $(\ell,m)$. The right panel displays the ES levels and their degeneracy according to Eq.~\eqref{eq:R}. The symmetry eigenvalue $s_i=\pm 1$ is denoted by blue and red, respectively. }
	\label{fig:es_levingu_12_pbc}
\end{figure} 

As required, each eigenvector of $\rho_{{{\rm{edge}}}}$ has a well-defined subsystem symmetry. We find that the corresponding quantum number is given by $s_i = (-1)^{\ell+m}$, as predicted from a field theory analysis~\cite{chen2012chiral,santos2014symmetryprotected}. In particular, all states generated by a given primary field inherit its symmetry properties. In Fig.~\ref{fig:es_levingu_12_pbc}, the symmetry eigenvalues $s_i=\pm 1$ are indicated in blue and red, respectively.

\subsection{Entanglement spectrum for $L$ nondivisible by 3}
Our numerical results for lengths $L$ that are not multiples of 3 follow a reproducible sequence different from Eq.~(\ref{eq:R}). Instead, they approximate the pattern
shown in Table~\ref{table:newCFT} (cf.~Appendix~\ref{app:numericalLevinGu}).
\begin{table}
	\caption{ES for the Levin-Gu model on a cylinder with circumference $L$ not divisible by 3 as described by Eq.~(\ref{phi3}). We display the degeneracy, quantum numbers ($\ell,m$), and symmetry eigenvalues $s_i=(-1)^{\ell+m}$ of the low-lying states. The first descendants are denoted by $(\ell,m)'$. This spectrum is obtained in the thermodynamic limit from our numerical results in Appendix~\ref{app:numericalLevinGu}.} \label{table:newCFT}
\begin{tabular}{|p{1cm}|p{1cm}|p{2.5cm}|p{1cm}|}
	\hline
	$\Delta_i$ & Deg.& $(\ell,m)$ & $s_i$ \\ 
	\hline
	0 & 1&(0,0) & 1\\ 
	1/6& 1&(1,0) & -1\\ 
	1/2 & 2&(0, $\pm$1) & -1\\ 
	2/3 & 2 &(1,$\pm$1)& 1\\ 
	5/6 & 1& (-1,0) & -1\\ 
	1 & 2 &(0,0)' & 1\\ 
	7/6 & 2&(1,0)' & -1\\ 
	4/3 & 3 &(2,0), (-1,$\pm$1) & 1\\ 
	3/2 & 4& (0,$\pm$1)' & -1\\ 
	5/3 & 4& \dots & 1\\ 
	11/6 & 4& & -1\\
	2 & 7& & 1 \\
	13/6 & 7& & -1\\
	7/3 & ?& & 1 \\
	\hline
\end{tabular}
\end{table}
This sequence is captured by the modified free boson spectrum
\be
\label{phi3}
\Delta(\ell,m)=\frac{(\ell-\phi)^2 - \phi^2 }{2}+\frac{m^2}{2} +{\rm{integers}},
\ee
with $\phi=1/3$, and the same symmetry resolution $s_i = (-1)^{\ell+m}$ as before. The parameter $\phi$ can be viewed as a flux threading the cylinder and affects the ground state wavefunction like a many-body Aharonov-Bohm effect~\cite{santos2014symmetryprotected}. (In Appendix.~\ref{se:cor_ha} we provide additional details that demonstrate this 3 periodicity using correlation functions)

\subsection{Velocity of the edge CFT}
\label{sec.velocity}
\begin{figure}[t]
	\centering
	\includegraphics[width=\linewidth]{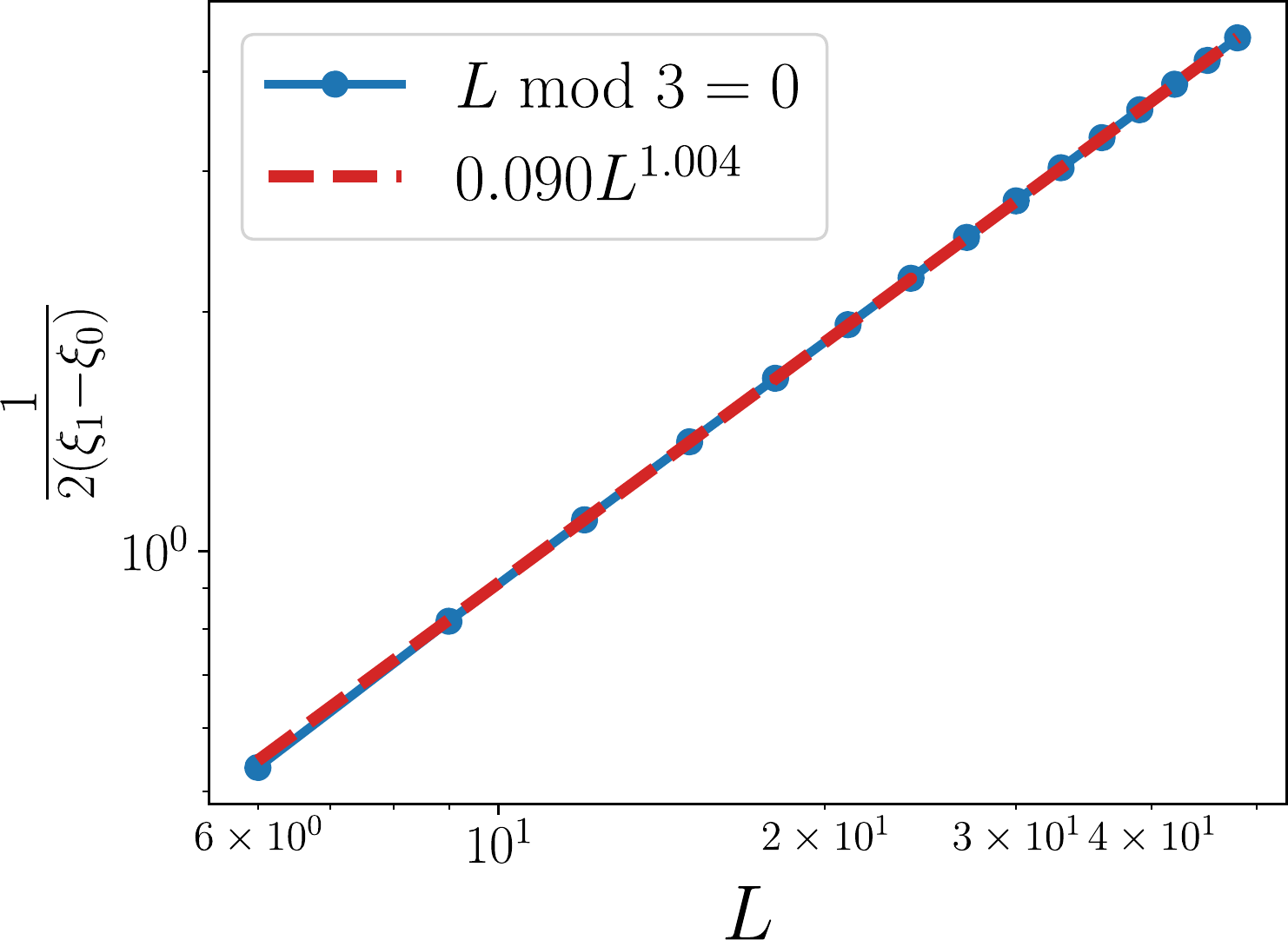}
		\includegraphics[width=\linewidth]{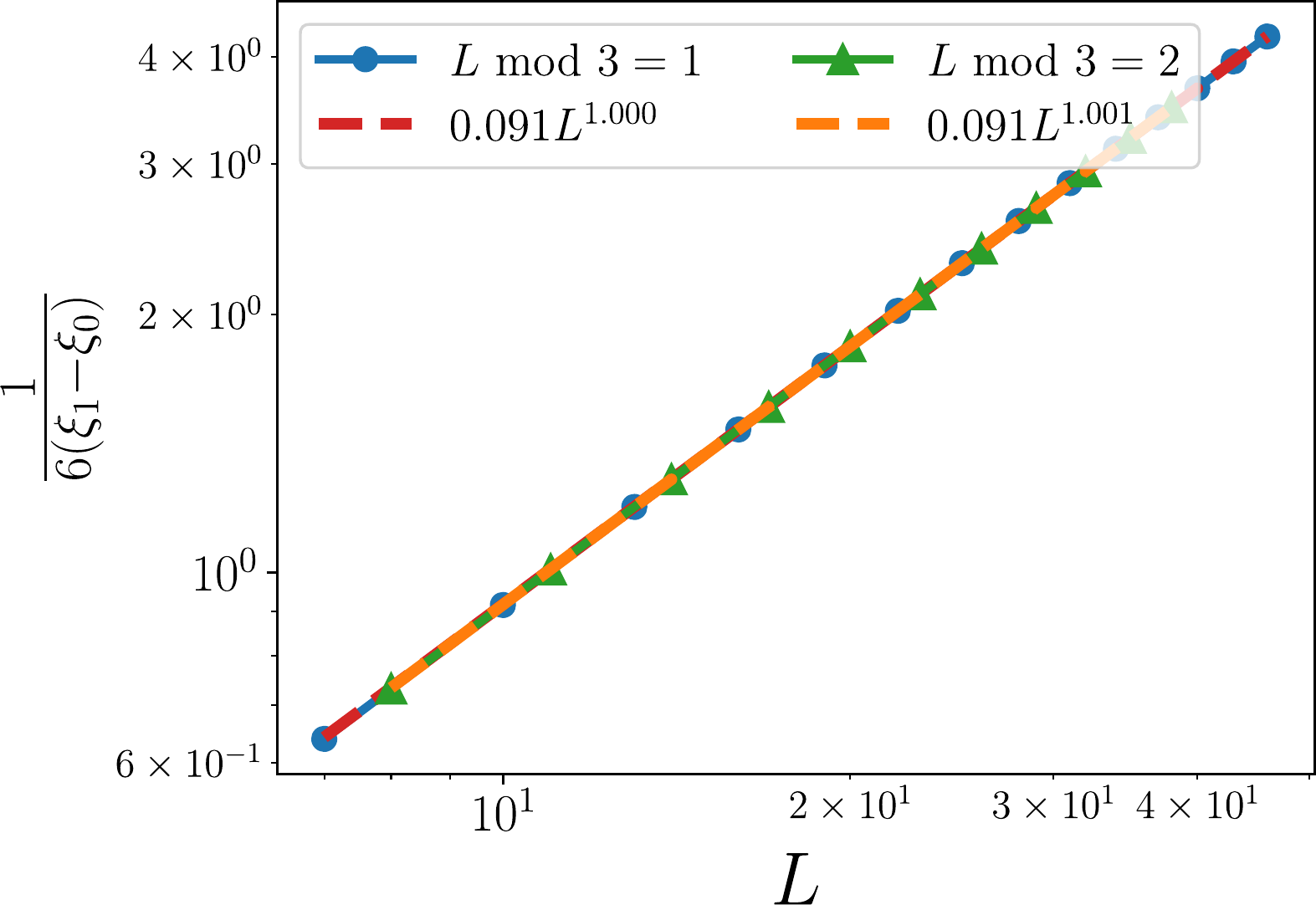}
	\caption{Inverse of the finite size gap in the quasienergy spectrum (i.e., eigenvalues of $- \log \rho_{{\rm{edge}}}$) as a function of system size $L$. From Eq.~(\ref{eq:voverL}), it allows to extract the velocity as in Eq.~(\ref{eq:v}), which is obtained both for $L$ divisible (upper panel) and not divisible (lower panel) by $3$.}
	\label{fig:es_eigenvalues_ratio}
\end{figure}

So far we determined the spectra $\{\Delta_i \}$ by matching numerical results to a CFT pattern of levels with unit spacing between descendants. This removes any ambiguity in the values of $\Delta_1$. We can now extract the velocity $v$ from the ES. Indeed, we have $\xi_1-\xi_0= \frac{v}{L} \Delta_1$. This gives 
\be
\label{eq:voverL}
\xi_1-\xi_0 =\begin{cases}
	\frac{ v}{2L}, & \text{for~L~divisible~by~3} \\
	\frac{ v}{6L}, & \text{for~L~not~divisible~by~3}.
\end{cases} 
\ee
These two cases are plotted in Fig.~\ref{fig:es_eigenvalues_ratio}. We thus obtain a value for the dimensionless velocity,
\be
\label{eq:v}
v \approx 11.023.
\ee

\subsection{Symmetry equidecomposition in the thermodynamic limit}
\begin{figure}[t]
	\centering
	\includegraphics[width=\linewidth]{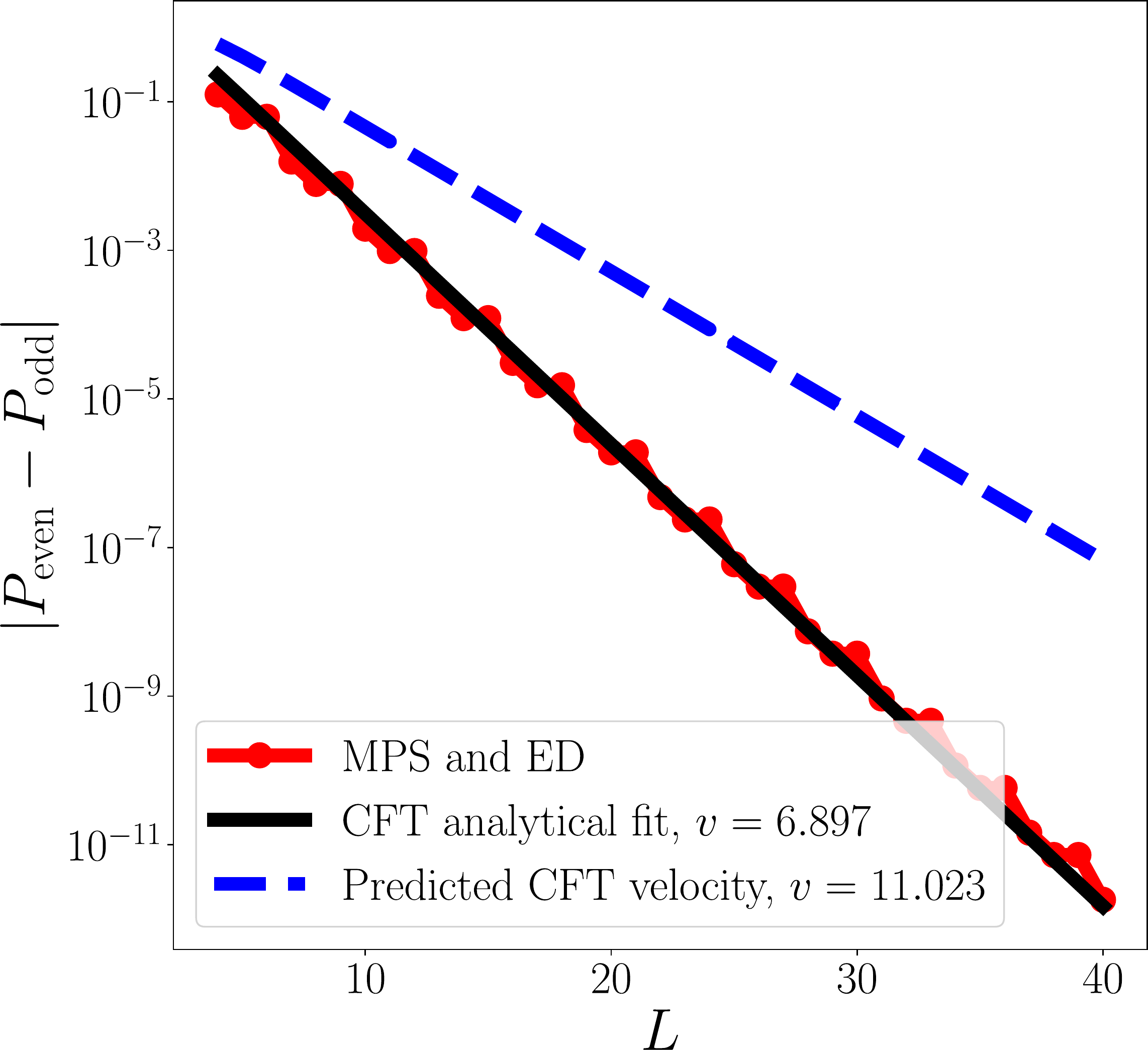}
	\caption{Difference between the parity-resolved probabilities obtained from our tensor network methods and confirmed using exact diagonalization for small systems, 
		as a function of system size. Upon increasing $L$ we obtain equipartition. We compare with the CFT prediction in Eq.~(\ref{eq:peven_podd}) for two values of $v$.}
	\label{fig:es_ratio}
\end{figure}

While the ES are different for the two symmetry sectors (blue vs.~red levels in Fig.~\ref{fig:es_levingu_12_pbc}), the total probabilities of the subsystem to be in either sector, 
\begin{align}
P_{ \rm{even}/\rm{odd}} =\frac{ \sum_{i \in \rm{even}/\rm{odd}} \lambda_i }{ \sum_{i } \lambda_i },
\end{align}
 converge quickly to 1/2 upon increasing $L$. The red curve in Fig.~\ref{fig:es_ratio} shows $|P_{\rm{even}}-P_{\rm{even}}|$, which decays exponentially with system size.

As an attempt at an analytic description of the decay of $P_{ \rm{even}} -P_{ \rm{odd}}$ with $L$, we employ the CFT spectrum, although the latter only describes the low-lying levels, whereas the probabilities $|P_{\rm{even}}-P_{\rm{even}}|$ presumably probe the entire spectrum. Nevertheless, from the CFT spectrum, we find
\bea
\label{eq:peven_podd}
P_{{\rm{even}}}-P_{{\rm{odd}}}&=&\frac{\sum_{\ell,m} (-1)^{\ell+m} e^{-\frac{ v}{L} (\ell^2+m^2)}}{\sum_{\ell,m} e^{-\frac{ v}{L} (\ell^2+m^2)}}\nonumber \\
&=& \left( \frac{\theta_4\left(e^{-\frac{ v}{L}}\right)}{\theta_3\left(e^{-\frac{ v}{L}}\right)} \right)^2\approx 
e^{-\frac{ L \pi^2}{2 v}},\,
\eea
where $\theta_i$ are Jacobi theta functions. In Fig.~\ref{fig:es_ratio}, we plot this difference for the velocity $v= 11.023$ (found in Sec.~\ref{sec.velocity}) as a dashed curve. The exponent does not quite match the numerical data, as expected. Curiously, the data can be fitted to Eq.~\ref{eq:peven_podd} with another value of the velocity, $v= 6.897$ (black curve).

\section{Deformed LG wavefunctions}
\label{se:MPS_cocycle} 
The Levin-Gu wavefunction studied in the previous section lies within the unique nontrivial 2D SPT phase protected by the $\intz_2 = \{1,-1\}$  symmetry. In this section, we would like to explore how the ES varies within the SPT phase. We start with a convenient but inessential simplification of the parent LG state. Keeping only the $CCZ$ gates in Eq.~(\ref{eq:LG}), we write
\be
\label{eq:LGprime}
|\Psi_\text{LG}' \rangle = U_{CCZ} |+ \rangle.
\ee
We study continuous deformations of this state given by
\begin{align}
\label{eq:psiLGprime}
 |\Psi_\text{LG}'(\lambda)\rangle_\text{cob} =U_\text{cob}(\lambda) |\Psi_\text{LG}'\rangle,
\end{align}
with
\begin{align}
 U_\text{cob}(\lambda) = \prod_{\text{triangles}} \left( \frac{\lambda(Z_j,Z_k,1)\lambda(Z_i,Z_j,1)}{\lambda(Z_i,Z_k,1)\lambda(Z_i,Z_j,Z_k)} \right)^{s_{ijk}}.
\end{align}
To define this transformation, one first enumerates all the vertices, and then denotes triangles by their ordered vertices $i<j<k$.  Here, $s_{ijk}=\pm 1$ is the orientation of the triangle~\cite{chen2013symmetry}. The $U(1)$-valued functions $\lambda$ are invariant under the global symmetry, i.e., $\lambda(Z_i,Z_j,Z_k)=\lambda(-Z_i,-Z_j,-Z_k)$. Due to this symmetry, $\lambda$ is parameterized by four independent phases. Each choice thereof yields a different wavefunction within the same nontrivial SPT phase since $ U_\text{cob}(\lambda)$ is a local symmetric unitary transformation. The resulting wavefunction is a special case of more general cocycle wavefunctions, which are reviewed in Appendix~\ref{app:MPS_cocycle}. In that context, $U_\text{cob}(\lambda)$ are referred to as coboundary transformations.

We can see that $U_\text{cob}(\lambda)$ can be incorporated into the quantum circuit form of Eq.~(\ref{eq:UAUBUAB}) with
\begin{align}
U_{A(B)} = \prod_{\text{triangles}\in {A(B)}} U_{CCZ}^{ijk} \times U^{A(B)}_\text{cob}(\lambda),
\end{align}
where
\be
 U^{A(B)}_\text{cob}(\lambda) = \prod_{\text{triangles} \in A(B)} \left(\frac{\lambda(Z_j,Z_k,1)\lambda(Z_i,Z_j,1)}{\lambda(Z_i,Z_k,1)\lambda(Z_i,Z_j,Z_k)} \right)^{s_{ijk}}.
\ee
Similarly, $U_{AB} = U_{AB}^{CCZ} U^{AB}_\text{cob}(\lambda)$, where
\be
\label{eq:uablambda}
 U^{AB}_\text{cob}(\lambda) = \prod_{\text{triangles} \in A \& B} \left( \frac{\lambda(Z_j,Z_k,1)\lambda(Z_i,Z_j,1)}{\lambda(Z_i,Z_k,1)\lambda(Z_i,Z_j,Z_k)} \right)^{s_{ijk}}.
\ee

\subsection{Coboundary transformations}
To explore how the ES is affected by coboundary transformations, we follow a specific path through the four-dimensional parameter space of $\lambda$. Specifically, we take $\lambda(1,1,1)=\lambda(-1,1,1)=e^{i \theta}$ and $\lambda(1,-1,1)=\lambda(-1,-1,1)=1$. This choice is arbitrary, and most other choices lead to the same phenomenology.
Next, we apply our tensor network methods to extract the ES of $|\Psi_\text{LG}'(\lambda)\rangle_\text{cob}$, together with its symmetry resolution. As derived in Appendix~\ref{app:MPS_cocycle}, the edge symmetry operator Eq.~(\ref{se:S}) is affected by the transformation.

\begin{figure}[t]
	\centering
	\includegraphics[width=\linewidth]{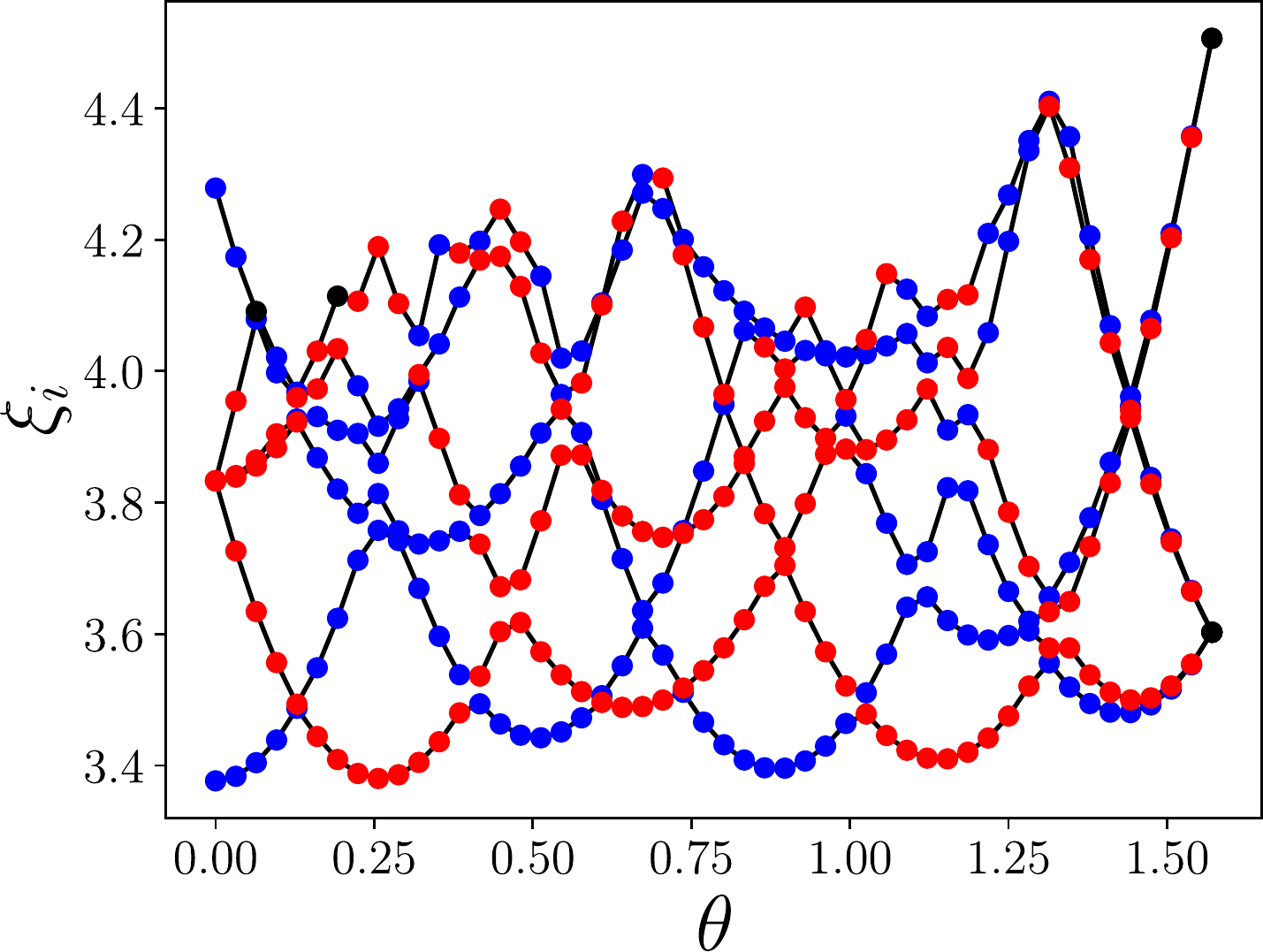}
	\caption{Parity-resolved ES for the deformed Levin-Gu wavefunction Eq.~(\ref{eq:psiLGprime}) upon varying the variable $\theta$ controlling the coboundary transformations. We show the first six levels as obtained from DMRG calculation for $L=12$.
		 The symmetry eigenvalue $s=\pm1$ is marked in blue and red, respectively. 
	}
	\label{fig:coboundary_es}
\end{figure}

\begin{figure}[t]
	\centering
	(a) \\
	\includegraphics[width=\linewidth]{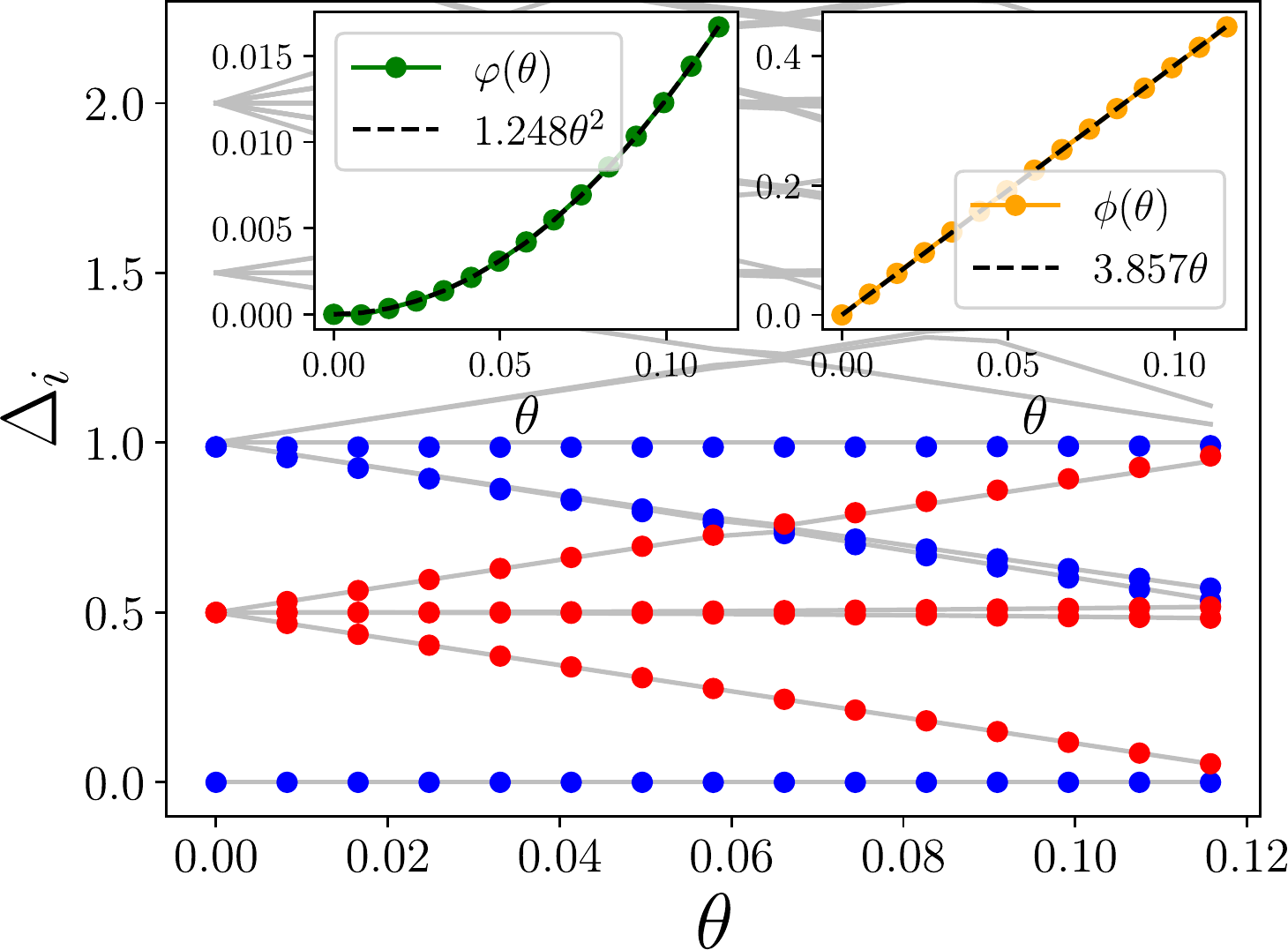}
	(b) \\
	\includegraphics[width=\linewidth]{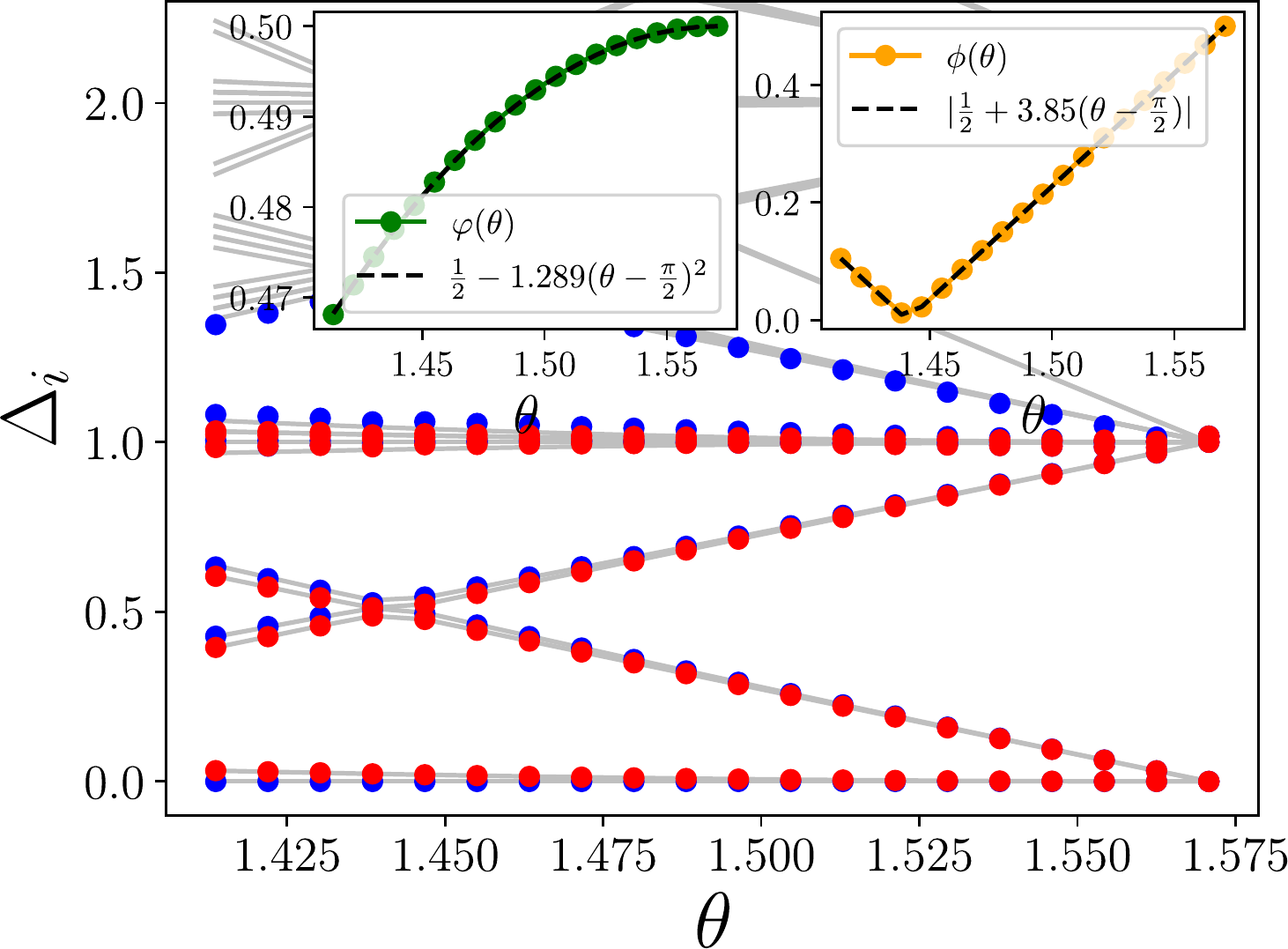}
	\caption{The same ES as in Fig.~\ref{fig:coboundary_es}, plotted in the form of the $\Delta_i$ defined in Eq.~(\ref{eq:scalingandshifting}). In (a) we focus on the vicinity of $\theta=0$, and in (b) we focus on the vicinity of $\theta=\pi/2$. The resulting levels match the CFT pattern Eq.~(\ref{eq:shiftedCFT}). $\phi(\theta), \varphi(\theta)$ are calculated for each point and plotted in each inset.
	At $\theta=\pi/2$, we have $\phi(\pi/2)=\varphi(\pi/2)=1/2$ corresponding to a 4-fold degenerate lowest level. See Appendix~\ref{app:fit_flux} regarding the generation of the fits.
	}
	\label{fig:coboundary_es_scaled}
\end{figure}

\subsection{Results}
We first construct the state $|+\rangle$, then apply $U_{AB}$ to get the MPS state on the zigzag chain $C_{AB}$. Although the bond dimension may change with coboundary transformation, we find that, in our case, it is constant at $\chi=9$. 
Subsequently, we construct $\rho_{{\rm{edge}}}$ as in Fig.~\ref{fig:schamatics2}(c), which doubles the bond due to the partial trace. Consequently, $\rho_{{\rm{edge}}}$ has the bond dimension $9^2$.

For each value of $\theta$, we obtain the low-lying eigenvalues and eigenvectors of the reduced density matrix. In Fig.~\ref{fig:coboundary_es}, we plot the symmetry-resolved quasienergies for the first excited levels. Our results show how the ES evolves with $\theta$. The obtained parabolic shapes motivate us to make an ansatz for the ES that corresponds to a flux insertion described by two additional parameters. Specifically, we expect the scaling dimensions [Eq.~(\ref{eq:scalingandshifting}) with a suitable choice of $v$]
to match the shifted CFT spectrum 
\bea
\label{eq:shiftedCFT}
\Delta(\ell,m) &=& \frac{(\ell-\phi(\theta))^2}{2}+\frac{(m-\varphi(\theta))^2}{2}+{\rm{integers}} \nonumber \\
&-& \frac{(\phi(\theta))^2}{2}-\frac{(\varphi(\theta))^2}{2},
\eea
where $\phi$ and $\varphi$ are the flux parameters. To assess whether the data follows this formula, we numerically fix $v,\phi$ and $\varphi$ for each $\theta$ from the first four energy levels, i.e., three excitations $\xi_i-\xi_0$. Having fixed all the parameters in our ansatz, we compute four additional levels and find that they are correctly predicted by Eq.~\eqref{eq:shiftedCFT}.
In Fig.~\ref{fig:coboundary_es_scaled}(a), we plot the scaled and shifted spectrum from the fit, for a small range of $\theta$ near $\theta=0$. We see that the CFT indeed changes from no flux to a finite flux that depends on $\phi(\theta), \varphi(\theta)$ as plotted in the inset. In Fig.~\ref{fig:coboundary_es_scaled}(b), we focus on the vicinity of the point $\theta=\pi/2$, which displays a degeneracy between the symmetry sectors, and also corresponds to the form of Eq.~\eqref{eq:shiftedCFT} with $\phi=\varphi=1/2$. These findings corroborate our suggestion that the coboundary transformation mediates a flux insertion.

We note that the compactification radius of the CFT that captures the ES remains constant, independent of our transformations. There is a priory no reason for a fixed radius. Instead, this behavior implies that our coboundary transformations describe a limited set of transformations within the SPT phase. On the other hand, our finding of the continuously varying fluxes already suggests a nontrivial structure of the SPT phase as a manifold.

\section{Wire decomposition of $H_{\rm{edge}}$}
\label{se:H1DLevin_Gu} 
In the previous sections we saw that the ES is sensitive to nonuniversal details such as the system size modulo an integer or coboundary transformations. Consequently, identifying the gapless edge theory from the ES involves guesswork, which may sometimes be difficult to achieve. In this section, we provide an algorithm to identify the gapless theory based on the gapped edge Hamiltonian $H_{{\rm{edge}}}$. The following decomposition 
is similar to field-theoretical wire constructions of 
fractional quantum Hall states~\cite{teo2014luttinger,james2013understanding,cano2015interactions,neupert2014wire}, SPTs~\cite{lu2012theory}, or other spin liquids~\cite{meng2015coupled,gorohovsky2015chiral,leviatan2020unification}.

We decompose the 1D edge Hamiltonian as 
\be
H_{{\rm{edge}}}=H_A+H_B+H_{AB},
\ee
where $H_{A(B)}$ acts only within $A(B)$, and $H_{AB}$ couples $A$ and $B$. Based on the Li-Haldane conjecture~\cite{li2008entanglement} it was argued~\cite{poilblanc2010entanglement,peschel2011relation,lauchli2012entanglement} that the ES between the legs of a two-leg ladder resembles the actual energy spectrum of each decoupled leg. Following this relationship in reverse, we suggest to infer the nature of the gapless theory describing the ES from the isolated chain $H_A$.

We remark that the study of $H_A$ reveals a number of interesting properties, such as a period $3$ modulation of correlation functions and a corresponding 3-fold dependence on $L$ of the finite size spectrum. Thus, the difference of the ES found in Sec.~\ref{se:Levin_Gu} depending on whether $L/3$ is a integer or not, has its origin already on the level of a single chain. Additionally, the model $H_A$ has the nice property that finite size properties converge very quickly to the infinite size limit. Below, we show results for the entanglement entropy of this 1D model.

\subsection{Spectrum of $H_A$ of the Levin-Gu model}
\label{se:LGHA}
\begin{figure}[t]
	\centering
	\includegraphics[width=\linewidth]{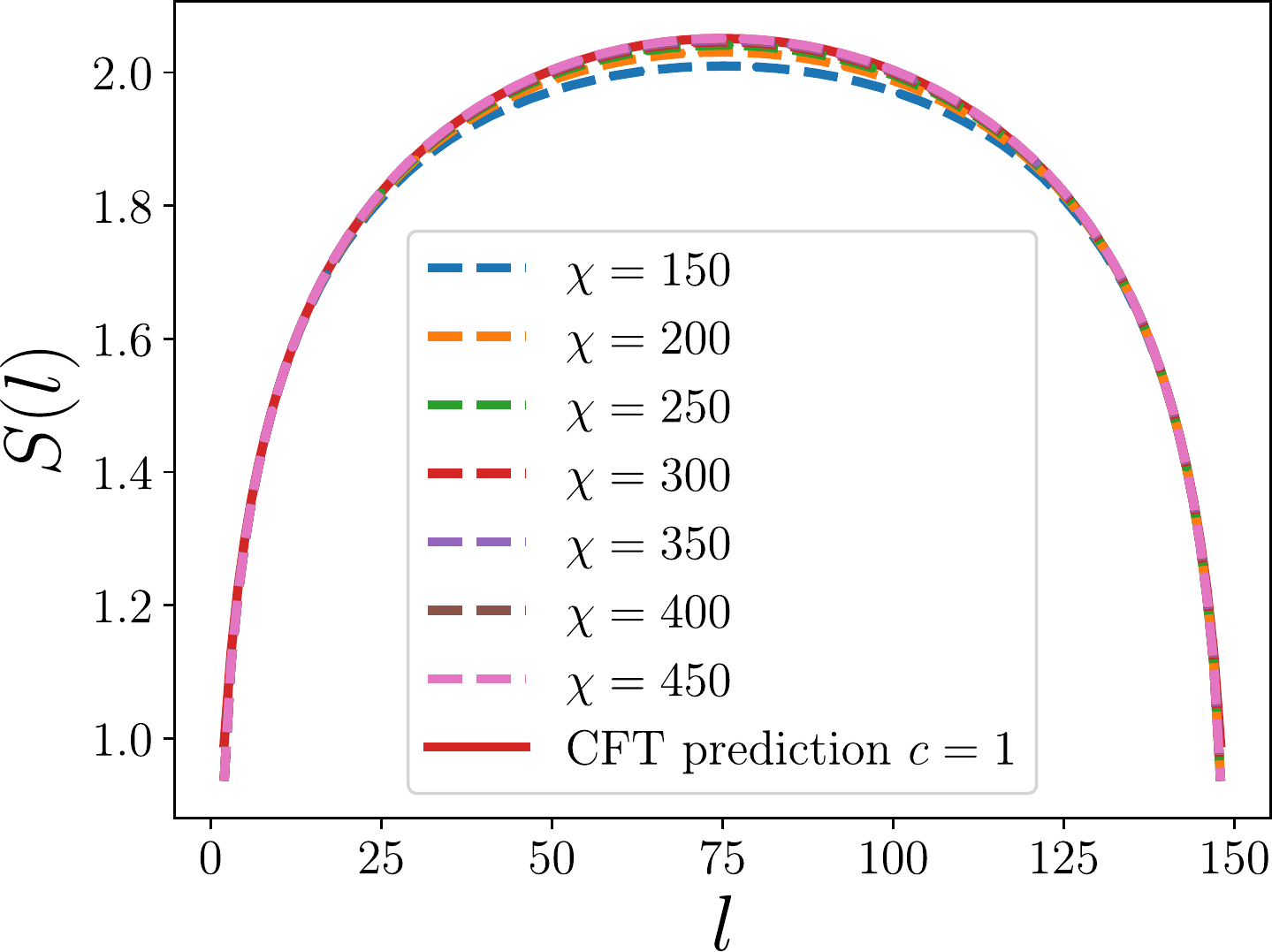}
	\caption{Entanglement entropy for the ground state of $H_A^{{\rm{LG}}}$ in Eq.~(\ref{eq:HLGA}) as a function of subsystem size $l$ obtained from a MPS with system size $L=150$ and varying bond dimension $\chi$.
	The numerical result converges to the CFT result Eq.~(\ref{eq:cftCC}) with central charge $c=1$.
	}		
	\label{fig:ent_spec}
\end{figure}

For the Levin-Gu model, the Hamiltonian $H_A^{LG}$ is given in Eq.~(\ref{eq:H1Dedge}), 
\be
\label{eq:HLGA}
H_A^\text{LG}=\sum_{i=1}^{L} [(Z_{i-1}X_iZ_{i+1}-X_i) +
(Z_{i-1}X_i + X_iZ_{i+1})], 
\ee
where here we only focus on subsystem $A$, so we label its sites by integer $i$ (rather than even integer). The decomposition of $H_A^\text{LG}$ into two terms is inessential for our purposes; it refers to the fact that, individually, each term is easily seen to yield a gapless spectrum. The fate of the full Hamiltonian is not readily apparent, and we determine it numerically. Specifically, by exploring the scaling of its ground state entanglement entropy, we show that $H_A^\text{LG}$ is gapless and extract the central charge $c$ of its field theory. In Fig.~\ref{fig:ent_spec}, we plot the entanglement entropy $S$ of a bipartition of this 1D chain of length $L$ as a function of subsystem size $l$. The numerical values are compared with the analytical CFT expression~\cite{clabrese2004entanglement}
\be
\label{eq:cftCC}
S(l) = \frac{c}{3} \log\left(\frac{L}{\pi} \sin\left[\frac{\pi l}{L}\right]\right) + \kappa,
\ee
where $c$ is the central charge and $\kappa$ is a nonuniversal constant. We find that $c=1$ agrees very well with the numerical data. 

\subsection{Robustness of the central charge upon coboundary transformations}
\begin{figure}[t]
	\centering
	\includegraphics[width=\linewidth]{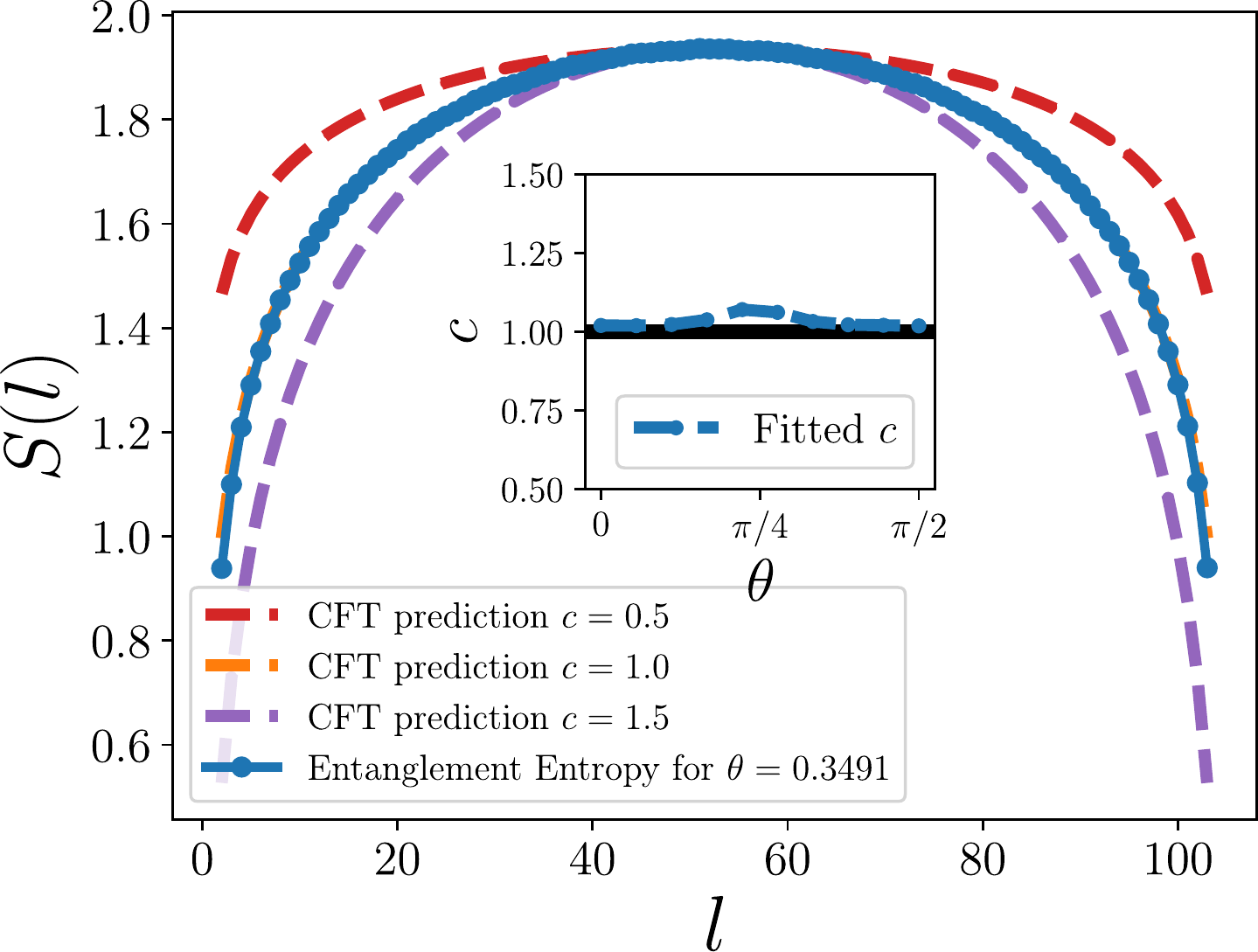}
	\caption{
	Entanglement entropy, as in Fig.~\ref{fig:ent_spec}, but for the ground state of Eq.~(\ref{eq:HAlambda}), describing the wire decomposition of the deformed Levin-Gu state, as parametrized by $\theta$, and also \emph{symmetry resolved} to a definite parity sector. In the main panel we consider the ground state of $H_A$ for a specific arbitrary point, $\theta \approx 0.3491$, and compare it with the CFT predictions for different central charges. We see that it matches $c=1$, as expected. In the inset, we plot the fit of the central charge for varying $\theta$, yielding $c \approx 1$ for all values of $\theta$ up to a maximal deviation of $\sim 7\%$. The fit is generated from $H^\theta_A$ of a subsystem of size $L=105$ and $\chi=4L=420$. }
	\label{fig:coboundary_central_charge}
\end{figure}

The central charge $c=1$ is expected to describe the edge theory for the entire SPT phase. We now consider the 1D edge Hamiltonian whose ground state is $|\Psi_\text{LG}'(\lambda)\rangle_\text{cob}$. Our aim is to show that $c$ is independent of $\lambda$. The explicit form of this Hamiltonian, and its decomposition as in Eq.~(\ref{eq:HABAN}), is cumbersome. In practice, we simply have
\be
\label{eq:HAlambda}
H_A(\lambda)={\rm{Tr}}_B (U_{AB} H_0 U_{AB}^\dagger),
\ee
where $U_{AB}$ depends on $\lambda$ as given in Eq.~(\ref{eq:uablambda}). Using our tensor network methods, we perform the partial trace over $B$ for the MPO representing $H=U_{AB} H_0 U_{AB}^\dagger$ [similar to Fig.~\ref{fig:schamatics2}(c)] to obtain an MPO representing $H_A$. We find its ground state and compute the entanglement entropy using DMRG. 

In Fig.~\ref{fig:coboundary_central_charge}, we show that coboundary transformations indeed preserve a good fit with $c=1$ for generic values of $\theta$, which parametrizes our coboundary transformation.

\section{Summary}
\label{se:summary}
1D SPTs display degeneracies in their ES~\cite{pollmann2010entanglement}. When performing a symmetry resolution of the ES, these degeneracies reflect an equidecomposition of the entanglement eigenvalues among different symmetry sectors~\cite{azses2020identification}.
Here, we showed that the gapless ES of 2D SPTs also have a natural symmetry decomposition. The gapless spectrum is described by a CFT and can be divided into towers of states which are descendants of primary fields. The symmetry quantum numbers are determined by the corresponding primary states. 

To demonstrate these claims explicitly for large systems on concrete models, we developed tensor-network-based methods. Starting from a $d$-dimensional tensor network of a short-range entangled state, we reduced the computation of the ES to a $d-1$-dimensional problem. For $d=2$, on which we focused, we ended up with effective 1D calculations that can be dealt with efficiently.

Our construction parallels the wire construction of topologically ordered phases, like the 2D fractional quantum Hall effect (FQHE). By breaking the 2D problem into coupled wires, the ES is universally stored in this inter-wire coupling~\cite{qi2012general}. Yet the wire construction of nonchiral SPT order~\cite{lu2012theory} has its unique features. In contrast to the wire construction of the chiral topological ordered case, in our nonchiral case we showed that the spectrum of the decoupled wires itself reflects the spectrum of the edge -- or equivalently~\cite{li2008entanglement} of the ES.

In the context of the FQHE and similar states, the wire construction allowed to explicitly construct effective quasi-1D synthetic realizations, e.g., using cold atoms ~\cite{mancini2015observation,petrescu2015chiral,cornfeld2015chiral,strinati2017laughlin,strinati2019pretopological}, which could be much easier to realize compared to their full 2D versions. In this sense, our 1D tensor network approach gives explicit quasi-1D models for 2D SPT order. This allows realizations of such states and explorations of their symmetry-resolved entanglement, e.g., on a small quantum computer, hence generalizing many existing realizations of 1D topological states, such as the cluster state~\cite{azses2020identification}.

	\section{Acknowledgments} We gratefully acknowledge support from the European Research Council (ERC) under the European Unions Horizon 2020 research and innovation programme under grant agreement
No. 951541, ARO (W911NF-20-1-0013) (ES), the US-Israel Binational
Science Foundation, grant number 2016255 (ES); the Israel Science Foundation, grant numbers 154/19 (ES) and 2572/21 (DFM) and the Deutsche Forschungsgemeinschaft (CRC/Transregio 183) (DFM). We thank Moshe Goldstein, Zohar Ringel, and Thomas Scaffidi for illuminating discussions.

\newpage
\newpage

\appendix

\section{Numerical results for the Levin-Gu model}
\label{app:numericalLevinGu}
In this appendix, we present numerical results for the ES of the Levin-Gu model calculated using the methods of Sections~\ref{se:general_construction} and ~\ref{se:Levin_Gu}. The lists of numerical results in Tables \ref{tab:mod3res} and \ref{tab:notmod3res} contain two columns for each $L$. The left column shows the bare eigenvalues $\lambda_i$ of the ES. Results for small systems confirmed by exact diagonalization are denoted by (ED). The second column of $\Delta_i$'s denoted ``CFT" is obtained from the first column by using Eq.~(\ref{eq:scalingandshifting}), by fixing $v$ such that $\Delta_1=\frac{1}{2}$ for $L$ being a multiple of $3$ ($L=12,18,24,30$), or $\Delta_1=\frac{1}{6}$ for $L$ not being a multiple of $3$ ($L=14,17,20$).

\begin{table}
\caption{\label{tab:mod3res}The largest eigenvalues of the ES for $L$ that is a multiple of $3$ obtained numerically mostly by using MPS methods described in the text. Blank lines separate degenerate eigenvalues to ease reading. The result of $L=12$ is a result of ED to provide a comparison of the two methods. The results are scaled in the CFT column to match the CFT spectrum described in the text.}
\begin{ruledtabular}
\begin{tabular}{|p{0.48\linewidth}|p{0.48\linewidth}|}
\begin{tabular}{p{0.48\linewidth} p{0.48\linewidth}}
	$L=12$ (ED) & CFT \\ 
0.03415& 0 \\\\
0.02164& 0.5\\
0.02164& 0.5\\
0.02164& 0.500004\\
0.02164& 0.500006\\\\
0.01386& 0.987772\\
0.01386& 0.987772\\
0.01386& 0.987774\\
0.01386& 0.987774\\
0.01386& 0.987774\\
0.01386& 0.987774\\\\
0.00861& 1.509386\\
\end{tabular}
& \begin{tabular}{p{0.48\linewidth} p{0.48\linewidth}}

	$L=18$ & CFT\\
0.005927& 0\\\\
0.004377& 0.5\\
0.004377& 0.500038\\
0.004376& 0.500224\\
0.004376& 0.500332\\\\
0.003241& 0.995632\\
0.003241& 0.995650\\
0.003241& 0.995566\\
0.003240& 0.996146\\
0.003241& 0.995580\\
0.003240& 0.996040\\\\
0.002376& 1.507274\\
\end{tabular}
\\
\hline
\begin{tabular}{p{0.48\linewidth} p{0.48\linewidth}}
	$L=24$ & CFT\\ 
0.00104116& 0\\\\
0.00082965& 0.5\\
0.00082933& 0.500836\\
0.00082957& 0.500206\\
0.00082866& 0.502614\\\\
0.00066105& 1.000170\\
0.00066145& 0.998830\\
0.00066110& 1.000012\\
0.00066109& 1.000022\\
0.00066051& 1.001982\\
0.00066053& 1.001900\\\\
0.00052246& 1.518194\\
\end{tabular}&
\begin{tabular}{p{0.48\linewidth} p{0.48\linewidth}}
	$L=30$ & CFT\\ 
0.00018325& 0\\\\
0.00015283& 0.5\\
0.00015283& 0.500048\\
0.00015234& 0.508910\\
0.00015230& 0.509600\\\\
0.00012711& 1.007720\\
0.00012668& 1.017110\\
0.00012707& 1.008628\\
0.00012720& 1.005916\\
0.00012646& 1.021908\\
0.00012647& 1.021750\\\\
0.00010430& 1.552650\\
\end{tabular}
\end{tabular}
\end{ruledtabular}
\end{table}

\begin{table}
\caption{\label{tab:notmod3res}The largest eigenvalues of the ES for $L$ that is not a multiple of $3$ obtained numerically mostly by using MPS methods described in the text. The result of $L=17$ is a result of ED to provide a comparison of the two methods. The results are scaled in the CFT column to match the CFT spectrum described in the text.}
\begin{ruledtabular}
\begin{tabular}{|p{0.48\linewidth}|p{0.48\linewidth}|}
\begin{tabular}{p{0.48\linewidth} p{0.48\linewidth}}
	$L=14$ & CFT\\ 
0.018205&0\\ 
0.015989&0.166666\\ 
0.012325&0.500964\\ 
0.012324&0.501028\\ 
0.010847&0.665022\\ 
0.010847&0.665024\\ 
0.009456&0.841195\\ 
0.00837&0.997875\\ 
0.008375&0.997212\\ 
0.007323&1.169509\\ 
0.007321&1.169927\\ 
0.006485&1.325660\\ 
0.006483&1.325973\\ 
0.006338&1.355135\\ 
0.005654&1.501629\\ 
0.005655&1.501508\\ 
\end{tabular} &
\begin{tabular}{p{0.48\linewidth} p{0.48\linewidth}}
	$L=17$ (ED) & CFT\\
0.007651817&0\\ 
0.006876625&0.166666\\ 
0.005552095&0.500502\\ 
0.005552092&0.500503\\ 
0.004995132&0.665447\\ 
0.004995132&0.665447\\ 
0.004470445&0.838606\\ 
0.004036603&0.997891\\ 
0.004036602&0.997891\\ 
0.003619472&1.168084\\ 
0.00361947&1.168085\\ 
0.003264707&1.329045\\ 
0.003264706&1.329045\\ 
0.003225258&1.348014\\ 
0.002925212&1.500374\\ 
0.002925211&1.500374\\ 
\end{tabular}
\\
\hline
\begin{tabular}{p{0.48\linewidth} p{0.48\linewidth}}
	$L=20$ & CFT\\ 
0.00321689&0\\ 
0.00293769&0.166666\\ 
0.00244834&0.501161\\ 
0.00244611&0.502827\\ 
0.00223668&0.667140\\ 
0.00223563&0.667999\\ 
0.00203912&0.836892\\ 
0.00185763&1.008016\\ 
0.0018588&1.006858\\ 
0.00169606&1.175047\\ 
0.0016941&1.177177\\ 
0.00155797&1.330952\\ 
0.00155589&1.333401\\ 
0.00154704&1.343873\\ 
0.00141089&1.512986\\ 
0.00141119&1.512589\\ 
\end{tabular} &
\begin{tabular}{p{0.48\linewidth} p{0.48\linewidth}}
$L=31$ & CFT\\
0.000134971& 0\\
0.000127297&0.166666\\
0.000113078&0.503877\\
0.000113039&0.504854\\
0.000106617&0.671371\\
0.000106650&0.670488\\
0.000100668&0.834815\\
0.0000941542&1.025287\\
0.0000942865&1.021286\\
0.0000887601&1.193246\\
0.0000888650&1.189886\\
0.0000843737&1.337537\\
0.0000843754&1.337478\\
0.0000843887&1.337029\\
0.0000789565&1.526459\\
0.0000788018&1.532042\\
\end{tabular}
\end{tabular}
\end{ruledtabular}
\end{table}

We can see that for $L$ being a multiple of 3, the shifted and rescaled spectrum is a good approximation of Eq.~(\ref{eq:R}), and for $L$'s that are not multiples of $3$ the ES follows the pattern in Eq.~(\ref{phi3}). 

\section{Exact diagonalization of the reduced density matrix for the Levin-Gu model}
\label{app:ExactDiagonalizationLevinGu}

\subsection{Eigenvalues of $\rho_A$}
Consider the ground state of Levin-Gu model $\ket{\psi}=\sum_{s_i\in \{0,1\}^S} U_{\mathrm{CCZ}} U_{\mathrm{CZ}} U_\mathrm{Z} \ket{s_i} $, where $S$ is the number of spins in the system and $U_{\mathrm{CCZ}}, U_{\mathrm{CZ}}, U_\mathrm{Z}$ are as defined in the main text. Here we focus on the reduced density matrix for subsystem $A$, which is obtained by partial trace over the complement of $A$, which we denote as $B$. Its eigenvalues give the ES of the system. We will show how we obtain the partial trace efficiently using numerical methods.

Let us now focus on the Hamiltonian dynamics of the resulting edge. The groundstate is 
\bea
|\Psi_\text{LG} \rangle &=& \sum_{s_A,s_B,s_{AB}} U_{\mathrm{CCZ}}(s_A,s_B,s_{AB}) U_{\mathrm{CZ}}(s_A,s_B,s_{AB}) \nonumber \\
&&~~~~~~~~U_\mathrm{Z}(s_A,s_B,s_{AB}) \ket{s_A,s_B,s_{AB}}.
\eea
As in the main text, by splitting the $U$'s pieces acting uniquely on $A,B$ or on the boundary $AB$, we are able to ignore and cancel the $A,B$ bulk parts, respectively. Hence, we have that the reduced density matrix entries are 
\bea \rho_{s_{A},s_{A}'} &=& \sum_{s_B} U_{\mathrm{CCZ}}^{\mathrm{AB}}(s_{A},s_B) U_{\mathrm{CCZ}}^{\mathrm{AB}}(s_{A}',s_B) \\ \nonumber &&~U^{\mathrm{AB}}_{\mathrm{CZ}}(s_{A },s_B) U^{\mathrm{AB}}_{\mathrm{CZ}}(s_{A}',s_B) 
,
\eea where $U^{\mathrm{AB}}$ are the unitaries that act on both subsystems $A$ and $B$, $s_{A,B}$ are binary strings representing the spins on the boundaries of $A, B$, and we use the standard ($Z$) basis of spins.

The structure of $\rho_A$ further decomposes into the product of matrix $M$ and its hermitian conjugate $M^\dagger$. Let's define $M =\sum_{s_A,s_B} U_{\mathrm{CCZ}}^{\mathrm{AB}}(s_{A},s_B) U^{\mathrm{AB}}_{\mathrm{CZ}}(s_{A},s_B) \ket{s_A}\bra{s_B}$ as the matrix representing the action of the unitaries on the edge qubits for a specific choice of $s_A$ and $s_B$. It is then clear that $\rho_A = M M^\dagger$, as the summation over the $B$ edge spins, is performed by the matrix multiplication, and the dagger matches the $B$ part to reproduce $\rho_A$. Hence, to obtain the eigenvalues of $\rho_A$, we only need to compute $M$, which is more efficient than computing $\rho_A$.

Moreover, we deduce the eigenvalues of $\rho_A$ from the matrix $M$. We verified numerically that $[M,M^\dagger]=0$ (we have yet to have an analytical proof for this). Hence, by the spectral theorem the eigenvalues of $\rho_A$ are simply the absolute value squared of the $M$ eigenvalues. The terms in $M$ are much easier to compute, and thus our method overall is about $2^L$ times faster than the Naive way of calculating $\rho_A$ as we do not sum over $s_B$, where $2L$ is the number of spins on the boundaries.

\subsection{Symmetry resolution of $\rho_A$}
In order to obtain the symmetry-resolved entanglement, one needs to calculate how the symmetry acts in the Gu-Levin basis. Gu-Levin basis $\ket{\sigma}$ is defined as the summation over all configurations in $A$ with a specific configuration $\sigma$ for the boundary spins, where each such configuration gets a sign $(-1)^{N_A}$ where $N_A$ is the number of domain walls calculated with $\ket{\uparrow}$ ghost spins on the boundary. We conjecture (the proof can be done by induction) that $X_A\ket{\sigma} = (-1)^{D+1} \ket{\bar{\sigma}}$ where $D$ is the number of domain walls on the boundary of $A$ divided by 2. Therefore, we see that the symmetry acts nontrivially on the boundary, as we expect for SPT phases.

To obtain the symmetry-resolved blocks of $\rho_A$, we apply the projection $\frac{I\pm X_A}{2}$ on both sides of $\rho_A$. This is done by computing $\rho_A = A A^\dagger$ and then computing $\frac{I\pm X_A}{2} \rho_A \frac{I\pm X_A}{2} = \frac{1}{4} \left[ \rho_A + X_A \rho_A X_A \pm \rho_A X_A \pm X_A \rho_A \right]$ element by element using the action of $X_A$ in the Levin-Gu basis. Similarly, one can get the momentum $k$ from the translation symmetry $T_A$ with similar projections. Therefore, the ES is constructed with its symmetry resolution using ED for systems up to length $L=12$.

\section{Correlations in $H_A$}
\label{se:cor_ha}

In this appendix, we further explore manifestations of the three-fold periodicity. In the main text we argued that the ES is described by a $c=1$ CFT for all system sizes $L$, but with different fluxes for $L=3n$ or $L \ne 3n$. Here, we explore the ground state properties of the Hamiltonian $H_A$ derived from Eq.~\ref{eq:HLGA} for the Levin-Gu model to understand the origin of this 3-fold periodicity. 

The model $H_A$ belongs to the family of Hamiltonians
\bea
H=- \left[ \sum_i \cos(\alpha)(X_i-Z_{i-1}X_iZ_{i+1}) \right. \nonumber \\
\left. \vphantom{\sum_i } +\sin(\alpha) (-Z_{i-1}X_i - X_iZ_{i+1}) \right] ,
\eea
which interpolates between gapless models. For $\alpha =0$ or $\alpha=\frac{\pi}{2}$, this Hamiltonian maps onto an XY model. Consequently, these cases are described by $c=1$ CFTs. The point $\alpha=\frac{\pi}{4}$ recovers Eq.~\ref{eq:HLGA}. It corresponds to the  Levin-Gu case, which is also a $c=1$ CFT, as we have seen in Sec.~\ref{se:LGHA}. We note that this 1D model has an interesting phase diagram as a function of $\alpha$ that may be worth exploring in detail. For our purposes, we now focus on the Levin-Gu case and show that the ground state of this Hamiltonian has period-three features in its correlation functions.

\begin{figure}[t]

	\includegraphics[width=\linewidth]{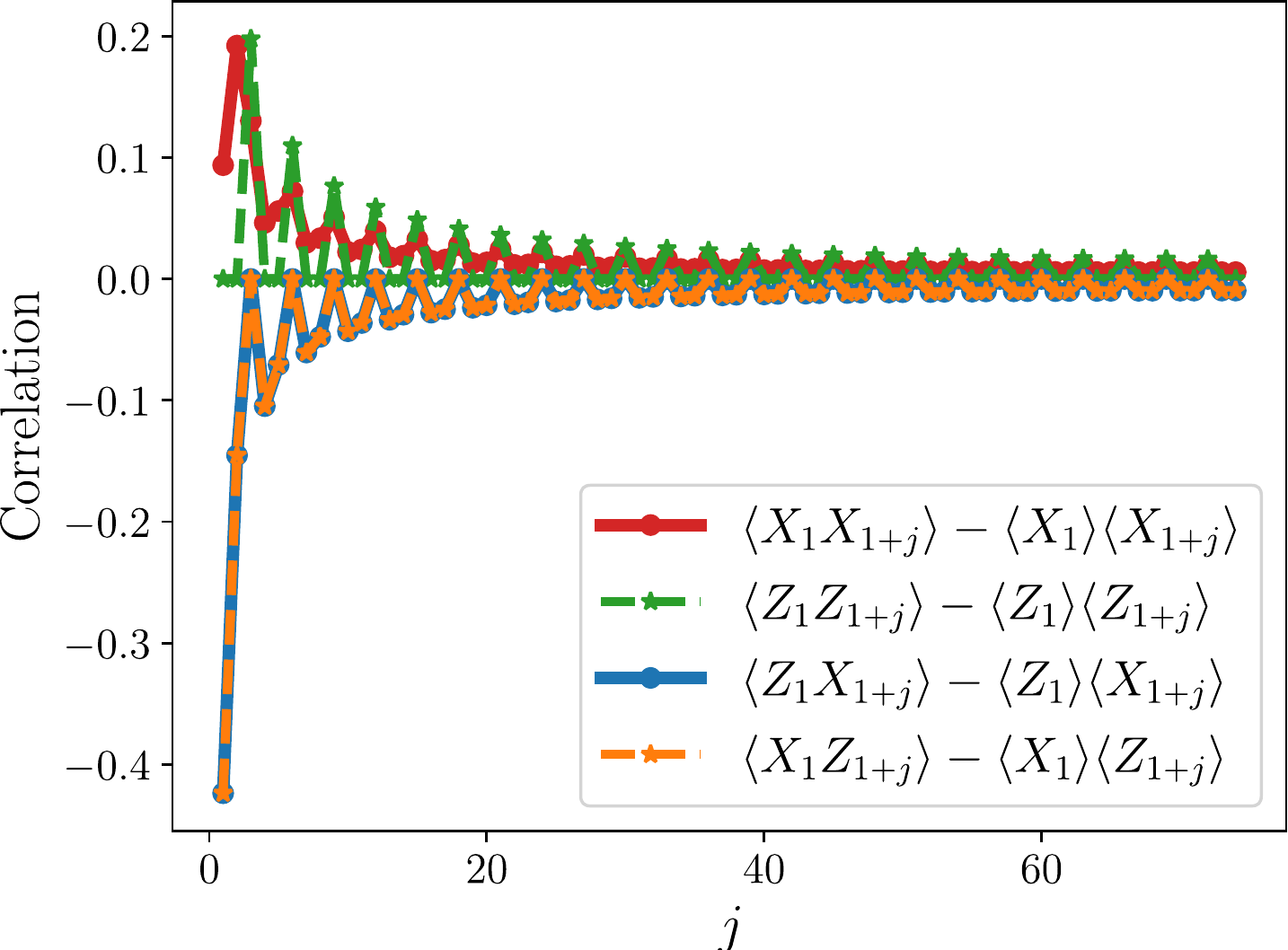}

 	\caption{\label{fig:corr_3points}Different correlation functions for the ground state of the Hamiltonian $H_A^{{\rm{LG}}}$ for  system size $L=150$. All these correlations exhibit three-fold periodicity.}
\end{figure}

\begin{figure*}
    \begin{minipage}[t]{0.48\linewidth}
    (a) \\
    \noindent\makebox[0.45\linewidth]{
	\includegraphics[width=\linewidth]{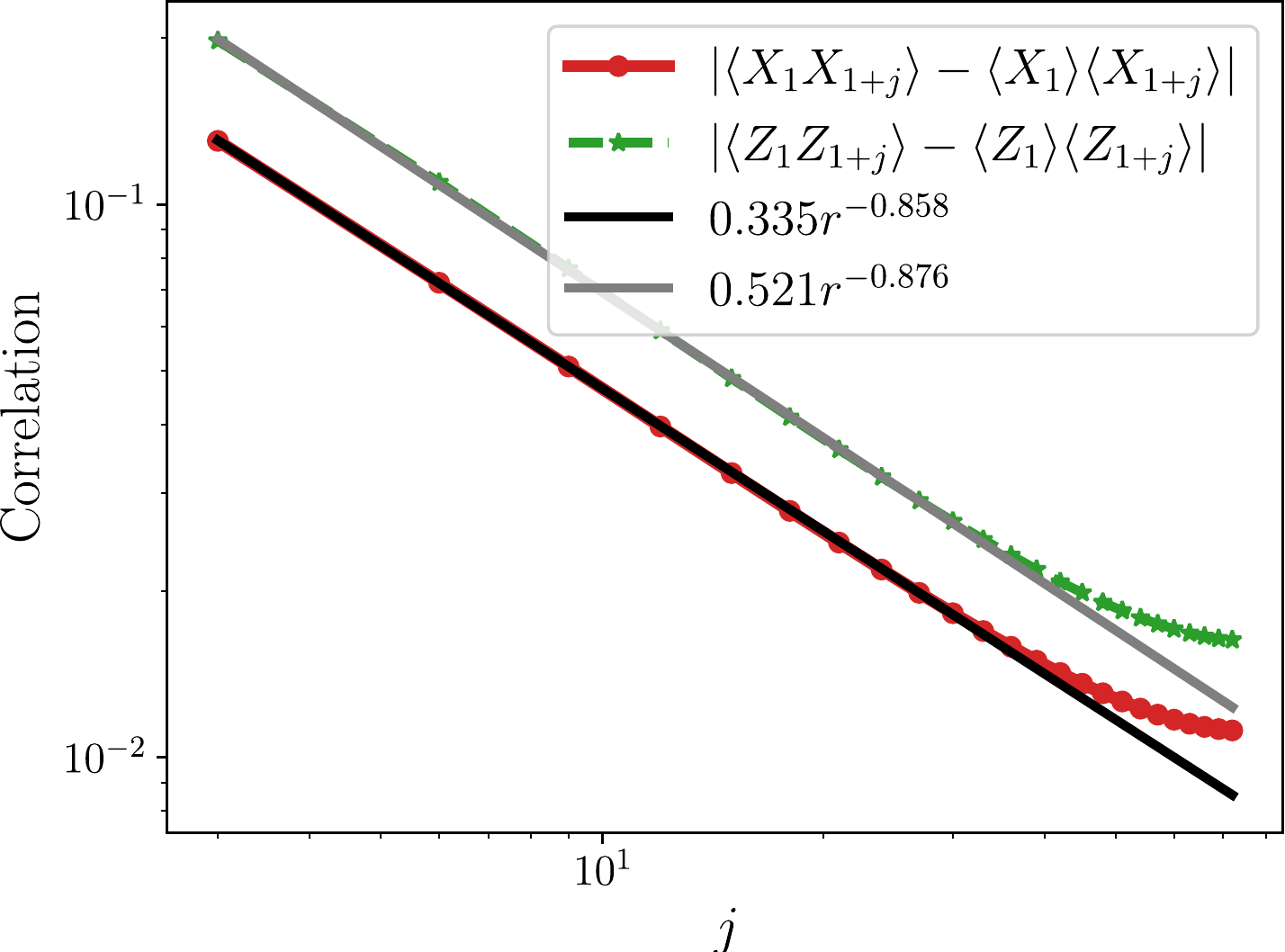}
    }
    \end{minipage} \hfill
    \begin{minipage}[t]{0.48\linewidth}
    (b) \\
    \noindent\makebox[0.45\linewidth]{
 	\includegraphics[width=\linewidth]{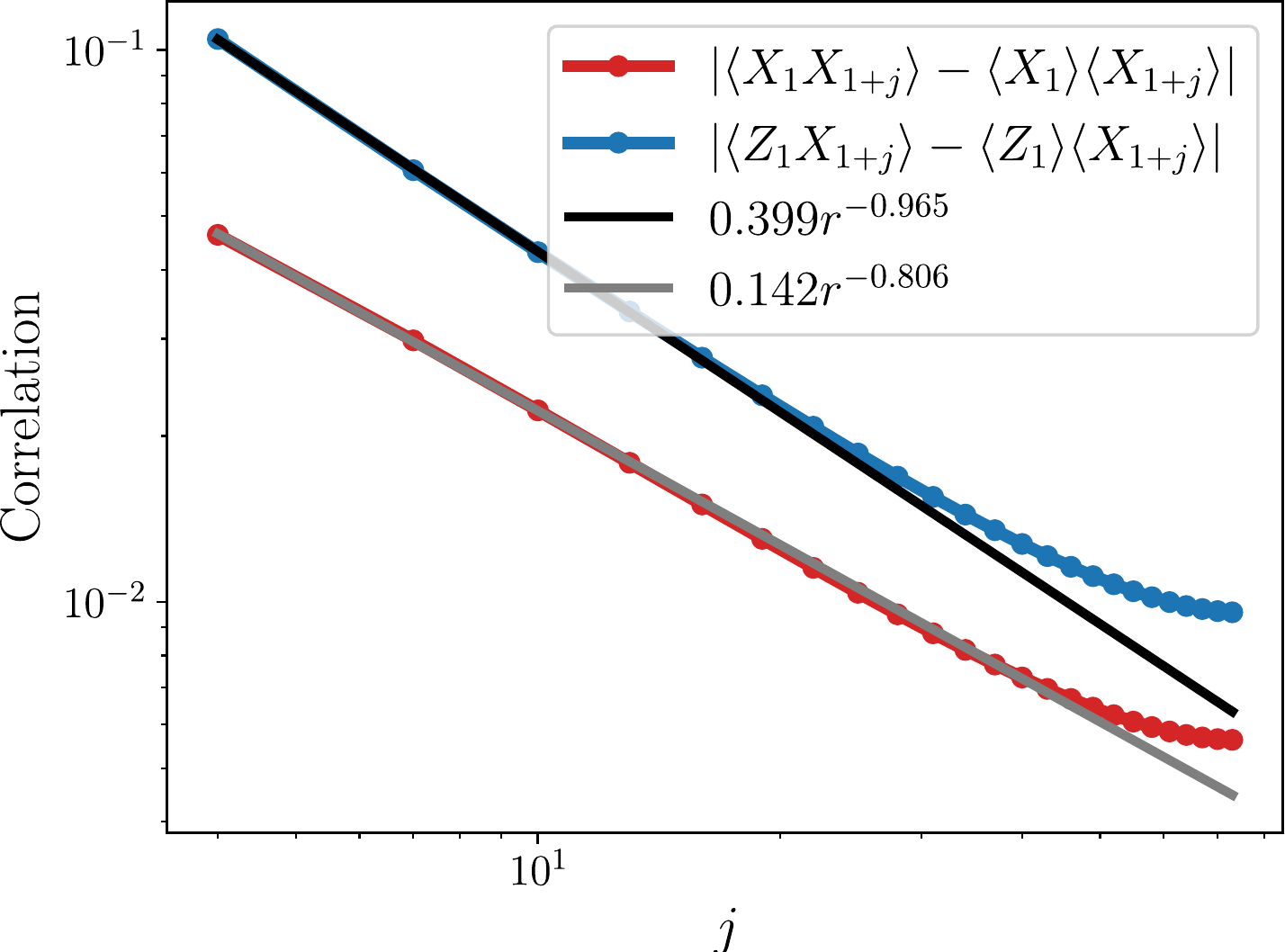}
  }
    \end{minipage} \hfill
  
    \begin{minipage}[t]{0.48\linewidth}
    (c) \\
  	\includegraphics[width=\linewidth]{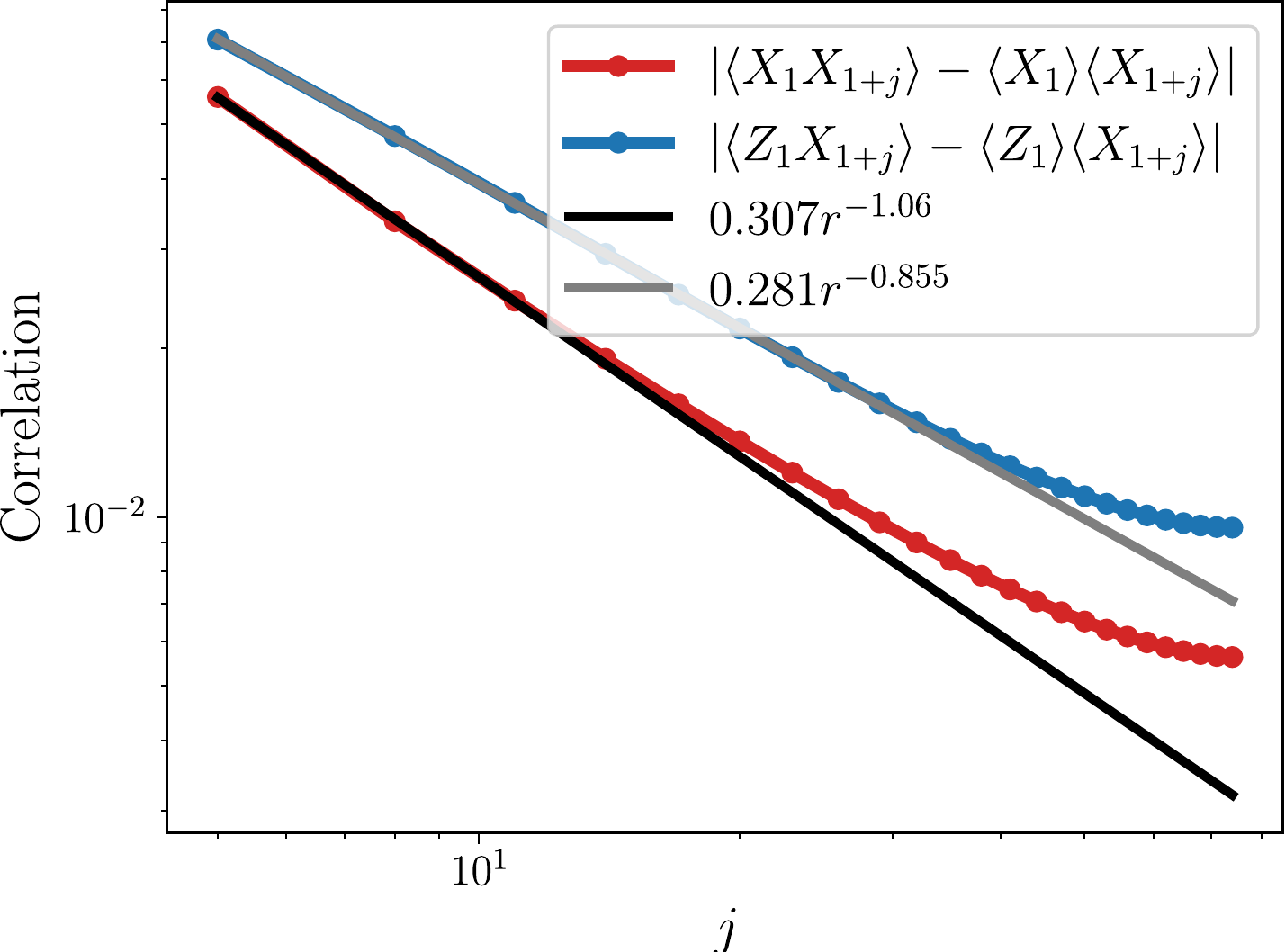}
    \end{minipage} \hfill
    
	\caption{\label{fig:corrfit_XZ+ZX_point} Power law fits for selected correlations on loglog scale for the Levin-Gu case for a system size $L=150$ and bond dimension $\chi=450$. The fits (a), (b), (c) are generated from the $j$ modulo $3$ $=$ $0$, $1$, $2$ points respectively. The correlations match power laws.}
\end{figure*}

We study the correlation functions $\expval{\sigma_i \sigma'_j} - \expval{\sigma'_i}\expval{\sigma'_j}$, where $\sigma,\sigma'=X,Y,Z$ are the different Pauli matrices. To obtain these correlations, we use the MPS procedure.  Refs.~\cite{orus2014practical,bridgeman2017hand,hastings2014notes} showed that the MPS always exhibits an exponential damping of the correlation functions, which may not approximate ground states with algebraic correlations (e.g. $|i-j|^{-\alpha}$). Hence, the MPS method introduces exponential errors in the large $|i-j|$ regime. These errors are circumvented by increasing the bond dimension $\chi$. Additionally, as we simulate only finite systems, we have finite-size effects.

In Fig.~\ref{fig:corr_3points} we show $\expval{\sigma_1 \sigma'_{1+j}} - \expval{\sigma'_1}\expval{\sigma'_{1+j}}$ as a function of $j$, for a system of size $L=150$ and bond dimension $\chi=450$. The three-fold periodicity is evident in all four correlation functions. This behavior explains the origin of the same periodicity seen in the ES in the main text as function of $L$.

To better understand these correlation functions, we separately analyze each of the three components with threefold periodicity. In Fig.~\ref{fig:corrfit_XZ+ZX_point}, we show them on a logarithmic scale. All non-zero correlation functions are consistent with a power-law decay. 
Some of the correlation function yields exponents close to unity, but others follow more unusual power laws with exponents $\approx 0.85$.

\section{Field theory}
\label{se:fieldtheory}
Refs.~\onlinecite{chen2012chiral,santos2014symmetryprotected} provide a field theory description of the physical edge of an SPT together with the action of the symmetry. We will accommodate the results of the symmetry-resolved entanglement within this field theory.

The field theory is constructed naturally by generalizing the $\intz_2$ symmetry to a $\intz_N$ symmetry. The edge theory consists of a pair of fields $\phi_{I}(x)$ ($I=1,2$), where $\frac{1}{2\pi} \partial_x \phi_2(x)$ is the canonical momentum conjugate to $\phi_1(x)$. The $\intz_N$ symmetry in the $p$-th phase $(p=0,1,\dots N-1)$ of $\mathcal{H}^3(\intz_N,U(1))=\intz_N$ is
\be
S_{{{\rm{edge}}}}^{(p)}=e^{\frac{i}{N} \left( \int_0^L dx \partial_x \phi_2+p \int_0^L dx \partial_x \phi_1 \right)}.
\ee 
We can recognize the product of two commuting factors, as in Eq.~(\ref{eq:Sedge}).

With the mode expansion
\be
\phi_I(x)=\phi_{0I}+K_{IJ}^{-1} P_{\phi J}\frac{2\pi}{L}x+i \sum_{n \ne 0} \frac{1}{n} \alpha_{I,n}e^{-inx\frac{2\pi}{L}},
\ee
where $[\phi_{0I}, P_{\phi J}]=i\delta_{IJ}$, $[\alpha_{I,n},\alpha_{J,m}]=nK^{-1}_{IJ} \delta_{n,-m}$, one obtains the canonical quantized fields with the commutation relations 
\be
[\phi_I(x_1),K_{I'J}\partial_x \phi_{J}(x_2)]=2 \pi i \delta_{I I'} \delta(x_1-x_2),
\ee
where $K_{IJ}=(\sigma^x)_{IJ}$. The winding numbers $(P_{\phi 1},P_{\phi 2})$ are integers denoted $(l,m)$ and determine the edge symmetry $S_{{{\rm{edge}}}}^{(p)}=e^{\frac{2 \pi i}{N} \left( P_{\phi 2} +p P_{\phi 1} \right)}$. 

Under inversion of the edge coordinate $\phi_1 \to \phi_1$ and $\phi_2 \to -\phi_2$, allowing us to define right- and left-moving fields $\phi_{R,L}=\phi_1 \pm \phi_2$. Based on this symmetry, the most general quadratic Hamiltonian can be written in terms of a pair of parameters as\begin{align}H =v \int \frac{dx}{(2 \pi)^2} \left( \frac{1}{R^2}(\partial_x \phi_1)^2 + \frac{R^2}{4}(\partial_x \phi_2)^2\right).\end{align} Substituting the mode expansion yields
\be
H=\frac{v}{L} \left( \left( \frac{\ell^2}{R^2}+ \frac{m^2 R^2}{4}\right) + \sum_{n>0} n(a_n^\dagger a_n+ b_n^\dagger b_n) \right)
\ee
where the bosonic creation and annihilation operators $a_n,b_n$ with canonical commutation relations are related to the $\alpha_{I,n}$ by a rescaling and Bogoliubov transformation. The primary edge states are labeled by $(\ell,m)$ and have the symmetry $S_{{{\rm{edge}}}}^{(p)}=e^{\frac{2 \pi i}{N} \left( m +p \ell \right)}$. One obtains infinite towers of states above these states generated by creating bosonic excitations. 

\section{2D cocycle wavefunctions}
\label{app:MPS_cocycle} 
In this appendix, we apply our tensor network methods for 2D cocycle wavefunctions. In contrast to the main text, we denote the group elements by $\intz_2 = \{0,1\}$, with `$0$' representing the identity, to match standard notation in the literature \cite{chen2013symmetry}. The Gu-Levin
$\intz_2$ wavefunction studied in the previous section lies within the unique nontrivial 2D
SPT phase. It corresponds to a specific cocycle wavefunction. Within the cohomological formalism, coboundary transformations allow us to move in the SPT phase and explore how the ES varies within the SPT phase.

\subsection{Cocycles wavefunctions}
We begin by briefly reviewing the construction of the cocycles wavefunction~\cite{chen2013symmetry}.
Consider a 2D triangular lattice on the 2-sphere. The Hilbert space at each site is spanned by the symmetry group elements $| g \rangle $, $g \in G=\intz_2 = \{0,1 \}$ (with an obvious generalization to $\intz_N$). The on-site action of the symmetry is $S_h|g\rangle = |hg \rangle$ with $h \in G$. The trivial symmetric ground state is given by $|+\rangle \equiv \otimes_i \left( \frac{1}{\sqrt{|G|}} \sum_g | g \rangle \right)$.

\begin{figure}[t]
	\centering
	\includegraphics[width=\linewidth]{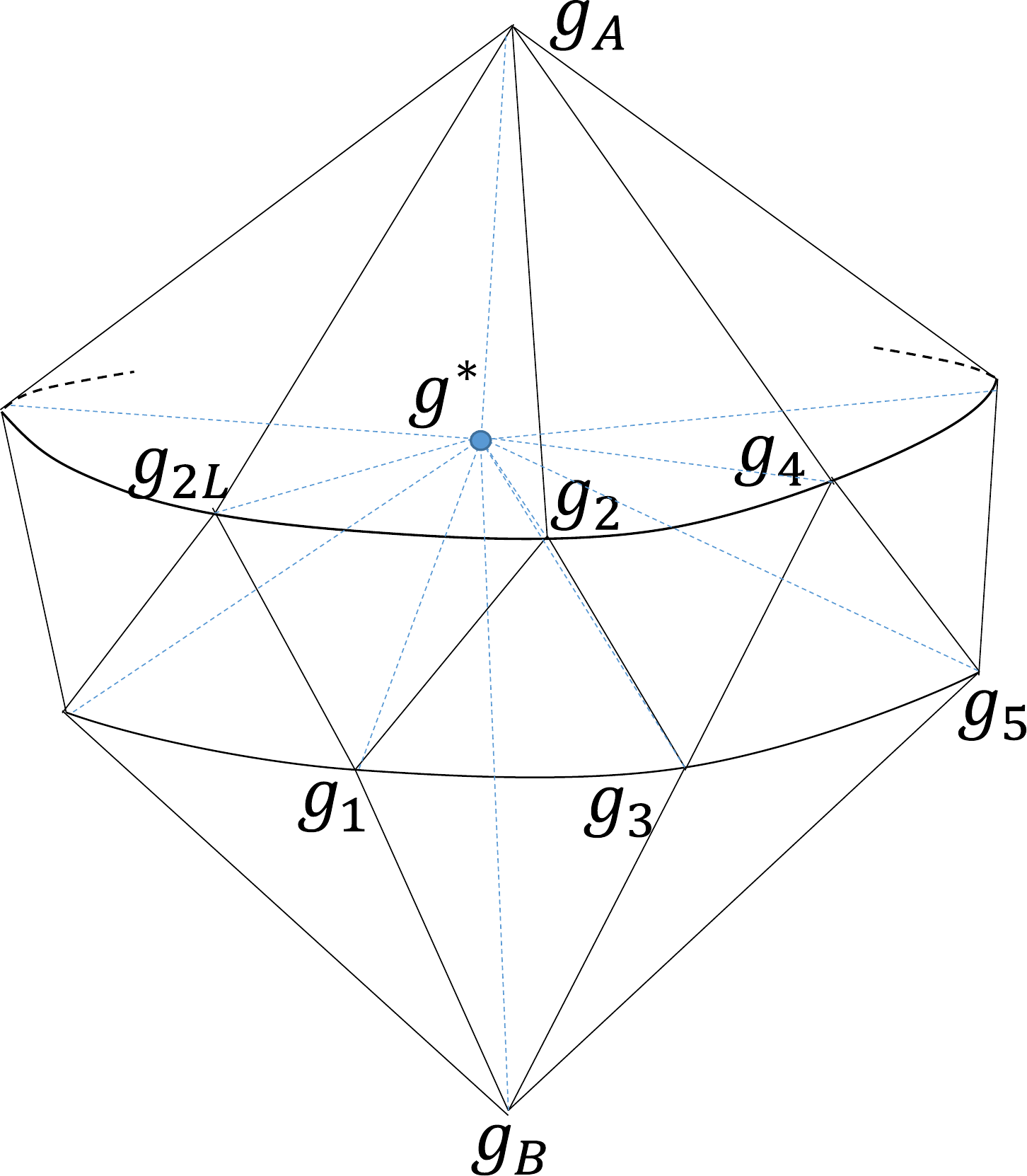}
	\caption{The triangulated 3D lattice used to construct the cocycle wavefunction. Region $A$ corresponds to the top part of the 2D surface, with one site in the ``bulk", $g_A$, and $L$ sites at the boundary. Similarly, subsystem $B$ contains only one site in the bulk, $g_B$. The 3D lattice has one site in its interior, denoted $g^*$.}
	\label{fig:diamond}
\end{figure}

The ideal SPT wavefunction is constructed as follows. We are interested in placing an SPT on the 2-sphere. In order to write its wavefunction we consider the 2-sphere as the boundary of a 3-ball. We minimally triangulate this $d+1$-dimensional space; to be specific we consider Fig.~\ref{fig:diamond} [see Fig.~\ref{fig:coboundary_es}(a) in Ref.~\onlinecite{azses2020symmetry}]. Here, we place only one site, $g^* =0$, in the interior of the 3-ball. The remaining sites are located on the 2-sphere, and the corresponding states are denoted $\{g_i \}$. Region $A$ consists of the upper hemisphere and includes $i={\rm{even}}=2,\dots,2L$ on the zigzag chain in Fig.~\ref{fig:schamatics}(c) and one extra site $g_A$ in the bulk of $A$. Similarly, region $B$ in the lower hemisphere includes the sites with odd $i$ and one extra site in the bulk of $B$ corresponding to a state $g_B$. 

The SPT state is written as $| \Psi \rangle = \sum_{\{g_i \}} \psi(\{g_i \})|\{g_i \} \rangle$, where the wavefunction $\psi(\{g_i \})$ is constructed using 3-cocycles, which reside on the tetrahedra, as explained next.

\subsection{Cochains, cocycles, and coboundaries}
3-cochains, $\nu_3(g_0,g_1,g_2,g_3)$, are $U(1)$-valued functions of $d+2=4$ $G$-valued variables $g_0,g_1,g_2,g_3$, $|\nu_3(g_0,g_1,g_2,g_3)|=1$, which satisfy $\nu_3(g_0,g_1,g_2,g_3)=\nu_3(h g_0,h g_1,h g_2,h g_3)$ for any $h \in G$.

3-Cocycles are special cochains satisfying $\prod_{i=0}^{4} \nu_3^{(-1)^i}(g_0,\dots,g_{i-1},g_{i+1},g_4)=1$, namely
\bea
\frac{\nu_3(g_1,g_2,g_3,g_4)\nu_3(g_0,g_1,g_3,g_4)\nu_3(g_0,g_1,g_2,g_3)}{\nu_3(g_0,g_2,g_3,g_4) \nu_3(g_0,g_1,g_2,g_4)}=1.\nonumber 
\eea

A 3-coboundary is a special 3-cocycle that is a `derivative' of a 2-cochain $\lambda_2(g_0,g_1,g_2)$,
\be
\label{eq:OURcocycle}
(d\lambda_2)(g_0,g_1,g_2,g_3)=\frac{\lambda_2(g_1,g_2,g_3)\lambda_2(g_0,g_1,g_3)}{\lambda_2(g_0,g_2,g_3) \lambda_2(g_0,g_1,g_2)}.
\ee
It automatically satisfies the cocycle condition. 

In our construction in Fig.~\ref{fig:diamond}, all tetrahedra consist of 4 vertices, containing $g_i,g_j,g_k$, which reside on the 2-sphere,  and $g^*$ which resides in the interior of the 3-ball. Each 3-cocycle $\nu_3$ corresponds to a wavefunction 
\be
\label{eq:cocycle}
\psi( \{ g_i \} ) = N \prod_{ijk \in \Delta} \nu_3^{s_{ijk}}(g_i,g_j,g_k,g^*),
\ee
where $g_i,g_j,g_k,g^*$ run over the vertices of all tetrahedra in Fig.~\ref{fig:diamond}. Here, ${s_{ijk}}=\pm 1$ is determined by the clockwise or anti-clockwise chirality of the triangle dictated by increasing order $i<j<k$. Here $g^*$ can be chosen arbitrarily~\cite{chen2013symmetry} to be $g^*=0$.

The cochain condition guarantees that $| \Psi \rangle =\sum_{\{ g_i\} } | \{ g_i\} \rangle$ is symmetric under the symmetry group $G$. Coboundary transformations of this state are like local unitary (LU) operators that transform one state into another within the same SPT phase. Two cocycles that belong to a different cohomology sector, as classified by $\mathcal{H}^3(G,U(1))$, are not connected by a finite depth circuit of LUs. The case of $G=\intz_2$ has exactly two sectors, $\mathcal{H}^3(G,U(1))=\intz_2$.

\subsection{Deformed Levin-Gu states}
We choose the following 3-cochain,
\be
\label{eq:coboundary_transformation}
\nu_3(g_i,g_j,g_k,0) = {\rm CCZ}(g_i,g_j,g_k) \frac{\lambda_2(g_j,g_k,0)\lambda_2(g_i,g_j,0)}{\lambda_2(g_i,g_k,0)\lambda_2(g_i,g_j,g_k)},
\ee
where $ {\rm CCZ}(g_i,g_j,g_k)=-1$ if $g_i=g_j=g_k=1$ and $ {\rm CCZ}(g_i,g_j,g_k)=1$ elsewhere. Here $\lambda_2$ is a 2-cochain parameterized by four independent phases $\lambda_2(g_i,g_j,0)$ ($i,j=0,1$).

The cocycle wavefunction Eq.~(\ref{eq:cocycle}) has the form of Eq.~(\ref{eq:UAUBUAB}) where $U_{A}$, $U_B$, and $U_{AB}$ are diagonal in the basis $|\{g_i \} \rangle$, $U_{A,B,AB}=\sum_{\{ g_i \} } U_{A,B,AB}(\{ g_i \}) | \{g_i \} \rangle \langle \{g_i \} |$ and given by
\bea
\label{eq:Ucocycle}
U_{AB}(\{ g_i \})&=& \frac{ \prod_{i=1}^{2L-2} \nu_3^{(-1)^i} (g_i,g_{i+1},g_{i+2},0)  }{\nu_3(g_1, g_{2L-1},g_{2L},0)} \nonumber \\
&\times & \nu_3(g_1,g_2,g_{2L},0),\nonumber \\
U_{A}(\{ g_i \})&=&\frac{ \prod_{i~\text{even}=2}^{2L-2} \nu_3(g_i,g_{i+2},g_A,0)}{\nu_3(g_2,g_{2L},g_A,0)},
\nonumber \\
U_{B}(\{ g_i \})&=&\frac{\prod_{i~\text{odd}=1}^{2L-3} \nu_3^{-1}(g_i,g_{i+2},g_B,0)}{\nu^{-1}_3(g_1,g_{2L-1},g_B,0)}.
\eea

\subsection{Symmetry operator}
The total symmetry operator 
\be
X=\sum_{ \{g_i \},g_A,g_B } |\{1-g_i \} ,1-g_A,1-g_B \rangle \langle \{ g_i \},g_A,g_B |,
\ee
separates into $X= X_A X_B$. The effective symmetry acting on $\rho_{{\rm{edge}}}$ is $U_A^\dagger X_A U_A$. Using Eqs.~(\ref{eq:Ucocycle}) and (\ref{eq:OURcocycle}), we see that $U_A$ factorizes into a product of CCZ gates familiar from the Levin-Gu model, and an additional factor $U^{\lambda_2}$ associated with the coboundary transformation,
\be
U_A = \prod_{ijk} U_{CCZ}^{ijk} \times U^{\lambda_2}.
\ee
 The latter factor, $U^{\lambda_2}$, consists of a product over many 2-cochains $\lambda_2$ acting on various triangles. It can be divided into two factors as $U^{\lambda_2} = U^{\lambda_2'} U^{\lambda_2''}$:
\begin{itemize}
	\item $U^{\lambda_2'}$: The 2-cocycles $\lambda(g_i,g_j,g_k)$ act only on triangles in $A$ within the 2-sphere. Due to the cochain condition $\lambda(g_i,g_j,g_k)=\lambda(1-g_i,1-g_j,1-g_k)$, we have ${U^{\lambda_2'}}^\dagger X_A U^{\lambda_2'}=X_A$.
	\item $U^{\lambda_2''}$: The 2-cocycles $\lambda(g_i,g_j,0)$ act on triangles on the surface defined by a membrane extending into the 3-ball from $\partial A$, 
		\be
	U^{\lambda_2''}= \sum_{\{g_i \}} \frac{\prod_{i ~even=2}^{2L-2} \lambda_2(g_i,g_{i+2},0)}{\lambda_2(g_{2},g_{2L},0) } | \{g_i \} \rangle \langle \{g_i \} |.
	\ee
	Since the symmetry operator $X$ does not act on the interior of the 3-ball, these are nontrivial. 
		\end{itemize}

As a result, we obtain
	\begin{align}
&	{	U^{\lambda_2''}}^\dagger X 	U^{\lambda_2''}=\\ &X \times 
	\sum_{\{g_i \}} \frac{\prod_{i ~even=2}^{2L-2} \frac{\lambda_2(g_i,g_{i+2},0)}{\lambda_2(1-g_i,1-g_{i+2},0)}}{\frac{\lambda_2(g_{2},g_{2L},0)}{ \lambda_2(1-g_{2},1-g_{2L},0)}} | \{g_i \} \rangle \langle \{g_i \} |. \nonumber
	\end{align}

As a result, the edge symmetry operator is 
\begin{align}
S_{{\rm{edge}}}&= \prod_{i \in even} X_i \prod_{i \in even} Z_i\prod_{i \in even} U_{CZ}^{i,i+2} \times \\
&\sum_{\{g_i \}} \frac{\prod_{i ~even=2}^{2L-2} \frac{\lambda_2(g_i,g_{i+2},0)}{\lambda_2(1-g_i,1-g_{i+2},0)}}{\frac{\lambda_2(g_{2},g_{2L},0)}{ \lambda_2(1-g_{2},1-g_{2L},0)}} | \{g_i \} \rangle \langle \{g_i \} |. \nonumber
\end{align}
This formula is used to determine the parity eigenvalue in Figs.~\ref{fig:coboundary_es} and \ref{fig:coboundary_es_scaled}.
\section{Flux parameters fit}
\label{app:fit_flux}
To fit the parameters $\phi(\theta),\varphi(\theta)$ to the ES, we solve a system of linear equations. After scaling and shifting the ES, it matches Eq.~(\ref{eq:shiftedCFT}). To find the scaling coefficients and $\phi(\theta),\varphi(\theta)$, we write linear equations for the scaled levels $a(\xi_i-\xi_0)$. To obtain $a, \phi,\varphi$, we construct explicit equations for the levels $(l,m) = (\pm 1,0), (0,\pm 1)$, which we denote $r_i$. By simple algebra, one gets linear equations for $a,\phi,\varphi$ by equating the scaled ES to the CFT spectrum such that $a(\xi_{r_1}-\xi_0) = 0.5(1-\phi)^2 - 0.5\phi^2 = 0.5-\phi$. Hence, we get four equations:
\be
\begin{split}
a = \frac{0.5-\phi}{\xi_{r_1}-\xi_0} = \frac{0.5+\phi}{\xi_{r_2}-\xi_0} = \frac{0.5-\varphi}{\xi_{r_3}-\xi_0} = \frac{0.5+\varphi}{\xi_{r_4}-\xi_0}.
\end{split}
\ee
Assuming no degeneracies exist, there are enough linear equations. We now write it in a matrix form:
\be
\begin{pmatrix}
	1 & \frac{1}{\xi_{r_1}-\xi_0} & 0 \\
	1 & \frac{-1}{\xi_{r_2}-\xi_0} & 0 \\
	1 & 0 & \frac{1}{\xi_{r_3}-\xi_0}
\end{pmatrix}
\begin{pmatrix}
	a \\
	\phi \\
	\varphi
\end{pmatrix}
=
\begin{pmatrix}
	\frac{1}{2(\xi_{r_1}-\xi_0)} \\
	\frac{1}{2(\xi_{r_2}-\xi_0)} \\
	\frac{1}{2(\xi_{r_3}-\xi_0)}
\end{pmatrix}.
\ee
Solving these equations, we get $a,\phi,\varphi$ for each $\theta$ value. 
These equations require a guess for the indices $r_1,r_2,r_3$, and one has the freedom to choose three of the four available equations.

Let us focus on the results in Fig.~\ref{fig:coboundary_es_scaled}. For $\theta$ near $0$ in Fig.~\ref{fig:coboundary_es_scaled}(a), we choose $r_1=1,r_3=2,r_4=3$ and solve their corresponding equations. Similarly, for $\theta$ near $\frac{\pi}{2}$ in Fig.~\ref{fig:coboundary_es_scaled}(b), we choose $r_1=2,r_2=4,r_3=1$. In both cases the results show a linear behavior for $\phi(\theta)$ and quadratic behavior for $\varphi(\theta)$.

$\phi,\varphi$ have symmetries due to the CFT spectrum. Specifically, $\phi \rightarrow -\phi$ and $\varphi \rightarrow -\varphi$ do not change the ES as changing $m, l$ accordingly $m\rightarrow -m, l\rightarrow -l$ preserves the ES. Similarly, $\phi \rightarrow \phi+1, \varphi \rightarrow \varphi+1$ still keeps the ES the same as $m\rightarrow m-1, l\rightarrow l-1$, respectively. These symmetries allow us to choose $0\leq \phi,\varphi<1$.


\begin{thebibliography}{51}%
	\makeatletter
	\providecommand \@ifxundefined [1]{%
		\@ifx{#1\undefined}
	}%
	\providecommand \@ifnum [1]{%
		\ifnum #1\expandafter \@firstoftwo
		\else \expandafter \@secondoftwo
		\fi
	}%
	\providecommand \@ifx [1]{%
		\ifx #1\expandafter \@firstoftwo
		\else \expandafter \@secondoftwo
		\fi
	}%
	\providecommand \natexlab [1]{#1}%
	\providecommand \enquote  [1]{``#1''}%
	\providecommand \bibnamefont  [1]{#1}%
	\providecommand \bibfnamefont [1]{#1}%
	\providecommand \citenamefont [1]{#1}%
	\providecommand \href@noop [0]{\@secondoftwo}%
	\providecommand \href [0]{\begingroup \@sanitize@url \@href}%
	\providecommand \@href[1]{\@@startlink{#1}\@@href}%
	\providecommand \@@href[1]{\endgroup#1\@@endlink}%
	\providecommand \@sanitize@url [0]{\catcode `\\12\catcode `\$12\catcode
		`\&12\catcode `\#12\catcode `\^12\catcode `\_12\catcode `\%12\relax}%
	\providecommand \@@startlink[1]{}%
	\providecommand \@@endlink[0]{}%
	\providecommand \url  [0]{\begingroup\@sanitize@url \@url }%
	\providecommand \@url [1]{\endgroup\@href {#1}{\urlprefix }}%
	\providecommand \urlprefix  [0]{URL }%
	\providecommand \Eprint [0]{\href }%
	\providecommand \doibase [0]{https://doi.org/}%
	\providecommand \selectlanguage [0]{\@gobble}%
	\providecommand \bibinfo  [0]{\@secondoftwo}%
	\providecommand \bibfield  [0]{\@secondoftwo}%
	\providecommand \translation [1]{[#1]}%
	\providecommand \BibitemOpen [0]{}%
	\providecommand \bibitemStop [0]{}%
	\providecommand \bibitemNoStop [0]{.\EOS\space}%
	\providecommand \EOS [0]{\spacefactor3000\relax}%
	\providecommand \BibitemShut  [1]{\csname bibitem#1\endcsname}%
	\let\auto@bib@innerbib\@empty
	\bibitem [{\citenamefont {Chen}\ and\ \citenamefont
		{Wen}(2012)}]{chen2012chiral}%
	\BibitemOpen
	\bibfield  {author} {\bibinfo {author} {\bibfnamefont {X.}~\bibnamefont
			{Chen}}\ and\ \bibinfo {author} {\bibfnamefont {X.-G.}\ \bibnamefont {Wen}},\
	}\bibfield  {title} {\bibinfo {title} {Chiral symmetry on the edge of {{2D}}
			symmetry protected topological phases},\ }\href
	{https://doi.org/10.1103/PhysRevB.86.235135} {\bibfield  {journal} {\bibinfo
			{journal} {Phys. Rev. B}\ }\textbf {\bibinfo {volume} {86}},\ \bibinfo
		{pages} {235135} (\bibinfo {year} {2012})}\BibitemShut {NoStop}%
	\bibitem [{\citenamefont {Levin}\ and\ \citenamefont
		{Gu}(2012)}]{levin2012braiding}%
	\BibitemOpen
	\bibfield  {author} {\bibinfo {author} {\bibfnamefont {M.}~\bibnamefont
			{Levin}}\ and\ \bibinfo {author} {\bibfnamefont {Z.-C.}\ \bibnamefont {Gu}},\
	}\bibfield  {title} {\bibinfo {title} {Braiding statistics approach to
			symmetry-protected topological phases},\ }\href
	{https://doi.org/10.1103/PhysRevB.86.115109} {\bibfield  {journal} {\bibinfo
			{journal} {Phys. Rev. B}\ }\textbf {\bibinfo {volume} {86}},\ \bibinfo
		{pages} {115109} (\bibinfo {year} {2012})}\BibitemShut {NoStop}%
	\bibitem [{\citenamefont {Santos}\ and\ \citenamefont
		{Wang}(2014)}]{santos2014symmetryprotected}%
	\BibitemOpen
	\bibfield  {author} {\bibinfo {author} {\bibfnamefont {L.~H.}\ \bibnamefont
			{Santos}}\ and\ \bibinfo {author} {\bibfnamefont {J.}~\bibnamefont {Wang}},\
	}\bibfield  {title} {\bibinfo {title} {Symmetry-protected many-body
			{{Aharonov}}-{{Bohm}} effect},\ }\href
	{https://doi.org/10.1103/PhysRevB.89.195122} {\bibfield  {journal} {\bibinfo
			{journal} {Phys. Rev. B}\ }\textbf {\bibinfo {volume} {89}},\ \bibinfo
		{pages} {195122} (\bibinfo {year} {2014})}\BibitemShut {NoStop}%
	\bibitem [{\citenamefont {Scaffidi}\ and\ \citenamefont
		{Ringel}(2016)}]{scaffidi2016wave}%
	\BibitemOpen
	\bibfield  {author} {\bibinfo {author} {\bibfnamefont {T.}~\bibnamefont
			{Scaffidi}}\ and\ \bibinfo {author} {\bibfnamefont {Z.}~\bibnamefont
			{Ringel}},\ }\bibfield  {title} {\bibinfo {title} {Wave functions of
			symmetry-protected topological phases from conformal field theories},\ }\href
	{https://doi.org/10.1103/PhysRevB.93.115105} {\bibfield  {journal} {\bibinfo
			{journal} {Phys. Rev. B}\ }\textbf {\bibinfo {volume} {93}},\ \bibinfo
		{pages} {115105} (\bibinfo {year} {2016})}\BibitemShut {NoStop}%
	\bibitem [{\citenamefont {Han}\ \emph {et~al.}(2017)\citenamefont {Han},
		\citenamefont {Tiwari}, \citenamefont {Hsieh},\ and\ \citenamefont
		{Ryu}}]{han2017boundary}%
	\BibitemOpen
	\bibfield  {author} {\bibinfo {author} {\bibfnamefont {B.}~\bibnamefont
			{Han}}, \bibinfo {author} {\bibfnamefont {A.}~\bibnamefont {Tiwari}},
		\bibinfo {author} {\bibfnamefont {C.-T.}\ \bibnamefont {Hsieh}},\ and\
		\bibinfo {author} {\bibfnamefont {S.}~\bibnamefont {Ryu}},\ }\bibfield
	{title} {\bibinfo {title} {Boundary conformal field theory and
			symmetry-protected topological phases in $2+1$ dimensions},\ }\href
	{https://doi.org/10.1103/PhysRevB.96.125105} {\bibfield  {journal} {\bibinfo
			{journal} {Phys. Rev. B}\ }\textbf {\bibinfo {volume} {96}},\ \bibinfo
		{pages} {125105} (\bibinfo {year} {2017})}\BibitemShut {NoStop}%
	\bibitem [{\citenamefont {Goldstein}\ and\ \citenamefont
		{Sela}(2018)}]{goldstein2018symmetry}%
	\BibitemOpen
	\bibfield  {author} {\bibinfo {author} {\bibfnamefont {M.}~\bibnamefont
			{Goldstein}}\ and\ \bibinfo {author} {\bibfnamefont {E.}~\bibnamefont
			{Sela}},\ }\bibfield  {title} {\bibinfo {title} {Symmetry-resolved
			entanglement in many-body systems},\ }\href
	{https://doi.org/10.1103/PhysRevLett.120.200602} {\bibfield  {journal}
		{\bibinfo  {journal} {Phys. Rev. Lett.}\ }\textbf {\bibinfo {volume} {120}},\
		\bibinfo {pages} {200602} (\bibinfo {year} {2018})}\BibitemShut {NoStop}%
	\bibitem [{\citenamefont {Xavier}\ \emph {et~al.}(2018)\citenamefont {Xavier},
		\citenamefont {Alcaraz},\ and\ \citenamefont
		{Sierra}}]{xavier2018equipartition}%
	\BibitemOpen
	\bibfield  {author} {\bibinfo {author} {\bibfnamefont {J.~C.}\ \bibnamefont
			{Xavier}}, \bibinfo {author} {\bibfnamefont {F.~C.}\ \bibnamefont
			{Alcaraz}},\ and\ \bibinfo {author} {\bibfnamefont {G.}~\bibnamefont
			{Sierra}},\ }\bibfield  {title} {\bibinfo {title} {Equipartition of the
			entanglement entropy},\ }\href {https://doi.org/10.1103/PhysRevB.98.041106}
	{\bibfield  {journal} {\bibinfo  {journal} {Phys. Rev. B}\ }\textbf {\bibinfo
			{volume} {98}},\ \bibinfo {pages} {041106} (\bibinfo {year}
		{2018})}\BibitemShut {NoStop}%
	\bibitem [{\citenamefont {Bonsignori}\ \emph {et~al.}(2019)\citenamefont
		{Bonsignori}, \citenamefont {Ruggiero},\ and\ \citenamefont
		{Calabrese}}]{bonsignori2019symmetry}%
	\BibitemOpen
	\bibfield  {author} {\bibinfo {author} {\bibfnamefont {R.}~\bibnamefont
			{Bonsignori}}, \bibinfo {author} {\bibfnamefont {P.}~\bibnamefont
			{Ruggiero}},\ and\ \bibinfo {author} {\bibfnamefont {P.}~\bibnamefont
			{Calabrese}},\ }\bibfield  {title} {\bibinfo {title} {Symmetry resolved
			entanglement in free fermionic systems},\ }\href
	{https://doi.org/10.1088/1751-8121/ab4b77} {\bibfield  {journal} {\bibinfo
			{journal} {Journal of Physics A: Mathematical and Theoretical}\ }\textbf
		{\bibinfo {volume} {52}},\ \bibinfo {pages} {475302} (\bibinfo {year}
		{2019})}\BibitemShut {NoStop}%
	\bibitem [{\citenamefont {Calabrese}\ \emph {et~al.}(2021)\citenamefont
		{Calabrese}, \citenamefont {Dubail},\ and\ \citenamefont
		{Murciano}}]{calabrese2021symmetry}%
	\BibitemOpen
	\bibfield  {author} {\bibinfo {author} {\bibfnamefont {P.}~\bibnamefont
			{Calabrese}}, \bibinfo {author} {\bibfnamefont {J.}~\bibnamefont {Dubail}},\
		and\ \bibinfo {author} {\bibfnamefont {S.}~\bibnamefont {Murciano}},\
	}\bibfield  {title} {\bibinfo {title} {Symmetry-resolved entanglement entropy
			in wess-zumino-witten models},\ }\href
	{https://doi.org/10.1007/jhep10(2021)067} {\bibfield  {journal} {\bibinfo
			{journal} {Journal of High Energy Physics}\ }\textbf {\bibinfo {volume}
			{2021}},\ \bibinfo {pages} {67} (\bibinfo {year} {2021})}\BibitemShut
	{NoStop}%
	\bibitem [{\citenamefont {Islam}\ \emph {et~al.}(2015)\citenamefont {Islam},
		\citenamefont {Ma}, \citenamefont {Preiss}, \citenamefont {Tai},
		\citenamefont {Lukin}, \citenamefont {Rispoli},\ and\ \citenamefont
		{Greiner}}]{islam2015measuring}%
	\BibitemOpen
	\bibfield  {author} {\bibinfo {author} {\bibfnamefont {R.}~\bibnamefont
			{Islam}}, \bibinfo {author} {\bibfnamefont {R.}~\bibnamefont {Ma}}, \bibinfo
		{author} {\bibfnamefont {P.~M.}\ \bibnamefont {Preiss}}, \bibinfo {author}
		{\bibfnamefont {M.~E.}\ \bibnamefont {Tai}}, \bibinfo {author} {\bibfnamefont
			{A.}~\bibnamefont {Lukin}}, \bibinfo {author} {\bibfnamefont
			{M.}~\bibnamefont {Rispoli}},\ and\ \bibinfo {author} {\bibfnamefont
			{M.}~\bibnamefont {Greiner}},\ }\bibfield  {title} {\bibinfo {title}
		{Measuring entanglement entropy in a quantum many-body system},\ }\href
	{https://doi.org/10.1038/nature15750} {\bibfield  {journal} {\bibinfo
			{journal} {Nature}\ }\textbf {\bibinfo {volume} {528}},\ \bibinfo {pages}
		{77} (\bibinfo {year} {2015})}\BibitemShut {NoStop}%
	\bibitem [{\citenamefont {Neven}\ \emph {et~al.}(2021)\citenamefont {Neven},
		\citenamefont {Carrasco}, \citenamefont {Vitale}, \citenamefont {Kokail},
		\citenamefont {Elben}, \citenamefont {Dalmonte}, \citenamefont {Calabrese},
		\citenamefont {Zoller}, \citenamefont {Vermersch}, \citenamefont {Kueng}
		\emph {et~al.}}]{neven2021symmetry}%
	\BibitemOpen
	\bibfield  {author} {\bibinfo {author} {\bibfnamefont {A.}~\bibnamefont
			{Neven}}, \bibinfo {author} {\bibfnamefont {J.}~\bibnamefont {Carrasco}},
		\bibinfo {author} {\bibfnamefont {V.}~\bibnamefont {Vitale}}, \bibinfo
		{author} {\bibfnamefont {C.}~\bibnamefont {Kokail}}, \bibinfo {author}
		{\bibfnamefont {A.}~\bibnamefont {Elben}}, \bibinfo {author} {\bibfnamefont
			{M.}~\bibnamefont {Dalmonte}}, \bibinfo {author} {\bibfnamefont
			{P.}~\bibnamefont {Calabrese}}, \bibinfo {author} {\bibfnamefont
			{P.}~\bibnamefont {Zoller}}, \bibinfo {author} {\bibfnamefont
			{B.}~\bibnamefont {Vermersch}}, \bibinfo {author} {\bibfnamefont
			{R.}~\bibnamefont {Kueng}}, \emph {et~al.},\ }\bibfield  {title} {\bibinfo
		{title} {Symmetry-resolved entanglement detection using partial transpose
			moments},\ }\href {https://doi.org/10.1038/s41534-021-00487-y} {\bibfield
		{journal} {\bibinfo  {journal} {npj Quantum Information}\ }\textbf {\bibinfo
			{volume} {7}},\ \bibinfo {pages} {152} (\bibinfo {year} {2021})}\BibitemShut
	{NoStop}%
	\bibitem [{\citenamefont {Vitale}\ \emph {et~al.}(2022)\citenamefont {Vitale},
		\citenamefont {Elben}, \citenamefont {Kueng}, \citenamefont {Neven},
		\citenamefont {Carrasco}, \citenamefont {Kraus}, \citenamefont {Zoller},
		\citenamefont {Calabrese}, \citenamefont {Vermersch},\ and\ \citenamefont
		{Dalmonte}}]{vitale2022symmetry}%
	\BibitemOpen
	\bibfield  {author} {\bibinfo {author} {\bibfnamefont {V.}~\bibnamefont
			{Vitale}}, \bibinfo {author} {\bibfnamefont {A.}~\bibnamefont {Elben}},
		\bibinfo {author} {\bibfnamefont {R.}~\bibnamefont {Kueng}}, \bibinfo
		{author} {\bibfnamefont {A.}~\bibnamefont {Neven}}, \bibinfo {author}
		{\bibfnamefont {J.}~\bibnamefont {Carrasco}}, \bibinfo {author}
		{\bibfnamefont {B.}~\bibnamefont {Kraus}}, \bibinfo {author} {\bibfnamefont
			{P.}~\bibnamefont {Zoller}}, \bibinfo {author} {\bibfnamefont
			{P.}~\bibnamefont {Calabrese}}, \bibinfo {author} {\bibfnamefont
			{B.}~\bibnamefont {Vermersch}},\ and\ \bibinfo {author} {\bibfnamefont
			{M.}~\bibnamefont {Dalmonte}},\ }\bibfield  {title} {\bibinfo {title}
		{{Symmetry-resolved dynamical purification in synthetic quantum matter}},\
	}\href {https://doi.org/10.21468/SciPostPhys.12.3.106} {\bibfield  {journal}
		{\bibinfo  {journal} {SciPost Phys.}\ }\textbf {\bibinfo {volume} {12}},\
		\bibinfo {pages} {106} (\bibinfo {year} {2022})}\BibitemShut {NoStop}%
	\bibitem [{\citenamefont {Rath}\ \emph {et~al.}(2022)\citenamefont {Rath},
		\citenamefont {Vitale}, \citenamefont {Murciano}, \citenamefont {Votto},
		\citenamefont {Dubail}, \citenamefont {Kueng}, \citenamefont {Branciard},
		\citenamefont {Calabrese},\ and\ \citenamefont
		{Vermersch}}]{rath2022entanglement}%
	\BibitemOpen
	\bibfield  {author} {\bibinfo {author} {\bibfnamefont {A.}~\bibnamefont
			{Rath}}, \bibinfo {author} {\bibfnamefont {V.}~\bibnamefont {Vitale}},
		\bibinfo {author} {\bibfnamefont {S.}~\bibnamefont {Murciano}}, \bibinfo
		{author} {\bibfnamefont {M.}~\bibnamefont {Votto}}, \bibinfo {author}
		{\bibfnamefont {J.}~\bibnamefont {Dubail}}, \bibinfo {author} {\bibfnamefont
			{R.}~\bibnamefont {Kueng}}, \bibinfo {author} {\bibfnamefont
			{C.}~\bibnamefont {Branciard}}, \bibinfo {author} {\bibfnamefont
			{P.}~\bibnamefont {Calabrese}},\ and\ \bibinfo {author} {\bibfnamefont
			{B.}~\bibnamefont {Vermersch}},\ }\bibfield  {title} {\bibinfo {title}
		{Entanglement barrier and its symmetry resolution: theory and experiment},\
	}\href@noop {} {\bibfield  {journal} {\bibinfo  {journal} {arXiv:2209.04393}\
		} (\bibinfo {year} {2022})}\BibitemShut {NoStop}%
	\bibitem [{\citenamefont {Azses}\ \emph {et~al.}(2020)\citenamefont {Azses},
		\citenamefont {Haenel}, \citenamefont {Naveh}, \citenamefont {Raussendorf},
		\citenamefont {Sela},\ and\ \citenamefont
		{Dalla~Torre}}]{azses2020identification}%
	\BibitemOpen
	\bibfield  {author} {\bibinfo {author} {\bibfnamefont {D.}~\bibnamefont
			{Azses}}, \bibinfo {author} {\bibfnamefont {R.}~\bibnamefont {Haenel}},
		\bibinfo {author} {\bibfnamefont {Y.}~\bibnamefont {Naveh}}, \bibinfo
		{author} {\bibfnamefont {R.}~\bibnamefont {Raussendorf}}, \bibinfo {author}
		{\bibfnamefont {E.}~\bibnamefont {Sela}},\ and\ \bibinfo {author}
		{\bibfnamefont {E.~G.}\ \bibnamefont {Dalla~Torre}},\ }\bibfield  {title}
	{\bibinfo {title} {Identification of symmetry-protected topological states on
			noisy quantum computers},\ }\href
	{https://doi.org/10.1103/PhysRevLett.125.120502} {\bibfield  {journal}
		{\bibinfo  {journal} {Phys. Rev. Lett.}\ }\textbf {\bibinfo {volume} {125}},\
		\bibinfo {pages} {120502} (\bibinfo {year} {2020})}\BibitemShut {NoStop}%
	\bibitem [{\citenamefont {Azses}\ \emph {et~al.}(2021)\citenamefont {Azses},
		\citenamefont {Dalla~Torre},\ and\ \citenamefont
		{Sela}}]{azses2021observing}%
	\BibitemOpen
	\bibfield  {author} {\bibinfo {author} {\bibfnamefont {D.}~\bibnamefont
			{Azses}}, \bibinfo {author} {\bibfnamefont {E.~G.}\ \bibnamefont
			{Dalla~Torre}},\ and\ \bibinfo {author} {\bibfnamefont {E.}~\bibnamefont
			{Sela}},\ }\bibfield  {title} {\bibinfo {title} {Observing floquet
			topological order by symmetry resolution},\ }\href
	{https://doi.org/10.1103/PhysRevB.104.L220301} {\bibfield  {journal}
		{\bibinfo  {journal} {Phys. Rev. B}\ }\textbf {\bibinfo {volume} {104}},\
		\bibinfo {pages} {L220301} (\bibinfo {year} {2021})}\BibitemShut {NoStop}%
	\bibitem [{\citenamefont {Azses}\ and\ \citenamefont
		{Sela}(2020)}]{azses2020symmetry}%
	\BibitemOpen
	\bibfield  {author} {\bibinfo {author} {\bibfnamefont {D.}~\bibnamefont
			{Azses}}\ and\ \bibinfo {author} {\bibfnamefont {E.}~\bibnamefont {Sela}},\
	}\bibfield  {title} {\bibinfo {title} {Symmetry-resolved entanglement in
			symmetry-protected topological phases},\ }\href
	{https://doi.org/10.1103/PhysRevB.102.235157} {\bibfield  {journal} {\bibinfo
			{journal} {Phys. Rev. B}\ }\textbf {\bibinfo {volume} {102}},\ \bibinfo
		{pages} {235157} (\bibinfo {year} {2020})}\BibitemShut {NoStop}%
	\bibitem [{\citenamefont {Pollmann}\ \emph {et~al.}(2010)\citenamefont
		{Pollmann}, \citenamefont {Turner}, \citenamefont {Berg},\ and\ \citenamefont
		{Oshikawa}}]{pollmann2010entanglement}%
	\BibitemOpen
	\bibfield  {author} {\bibinfo {author} {\bibfnamefont {F.}~\bibnamefont
			{Pollmann}}, \bibinfo {author} {\bibfnamefont {A.~M.}\ \bibnamefont
			{Turner}}, \bibinfo {author} {\bibfnamefont {E.}~\bibnamefont {Berg}},\ and\
		\bibinfo {author} {\bibfnamefont {M.}~\bibnamefont {Oshikawa}},\ }\bibfield
	{title} {\bibinfo {title} {Entanglement spectrum of a topological phase in
			one dimension},\ }\href {https://doi.org/10.1103/PhysRevB.81.064439}
	{\bibfield  {journal} {\bibinfo  {journal} {Phys. Rev. B}\ }\textbf {\bibinfo
			{volume} {81}},\ \bibinfo {pages} {064439} (\bibinfo {year}
		{2010})}\BibitemShut {NoStop}%
	\bibitem [{\citenamefont {Cornfeld}\ \emph {et~al.}(2019)\citenamefont
		{Cornfeld}, \citenamefont {Landau}, \citenamefont {Shtengel},\ and\
		\citenamefont {Sela}}]{cornfeld2019entanglement}%
	\BibitemOpen
	\bibfield  {author} {\bibinfo {author} {\bibfnamefont {E.}~\bibnamefont
			{Cornfeld}}, \bibinfo {author} {\bibfnamefont {L.~A.}\ \bibnamefont
			{Landau}}, \bibinfo {author} {\bibfnamefont {K.}~\bibnamefont {Shtengel}},\
		and\ \bibinfo {author} {\bibfnamefont {E.}~\bibnamefont {Sela}},\ }\bibfield
	{title} {\bibinfo {title} {Entanglement spectroscopy of non-abelian anyons:
			Reading off quantum dimensions of individual anyons},\ }\href
	{https://doi.org/10.1103/PhysRevB.99.115429} {\bibfield  {journal} {\bibinfo
			{journal} {Phys. Rev. B}\ }\textbf {\bibinfo {volume} {99}},\ \bibinfo
		{pages} {115429} (\bibinfo {year} {2019})}\BibitemShut {NoStop}%
	\bibitem [{\citenamefont {Else}\ \emph {et~al.}(2012)\citenamefont {Else},
		\citenamefont {Schwarz}, \citenamefont {Bartlett},\ and\ \citenamefont
		{Doherty}}]{else2012symmetry}%
	\BibitemOpen
	\bibfield  {author} {\bibinfo {author} {\bibfnamefont {D.~V.}\ \bibnamefont
			{Else}}, \bibinfo {author} {\bibfnamefont {I.}~\bibnamefont {Schwarz}},
		\bibinfo {author} {\bibfnamefont {S.~D.}\ \bibnamefont {Bartlett}},\ and\
		\bibinfo {author} {\bibfnamefont {A.~C.}\ \bibnamefont {Doherty}},\
	}\bibfield  {title} {\bibinfo {title} {Symmetry-protected phases for
			measurement-based quantum computation},\ }\href
	{https://doi.org/10.1103/PhysRevLett.108.240505} {\bibfield  {journal}
		{\bibinfo  {journal} {Phys. Rev. Lett.}\ }\textbf {\bibinfo {volume} {108}},\
		\bibinfo {pages} {240505} (\bibinfo {year} {2012})}\BibitemShut {NoStop}%
	\bibitem [{\citenamefont {Raussendorf}\ and\ \citenamefont
		{Briegel}(2001)}]{raussendorf2001one}%
	\BibitemOpen
	\bibfield  {author} {\bibinfo {author} {\bibfnamefont {R.}~\bibnamefont
			{Raussendorf}}\ and\ \bibinfo {author} {\bibfnamefont {H.~J.}\ \bibnamefont
			{Briegel}},\ }\bibfield  {title} {\bibinfo {title} {A one-way quantum
			computer},\ }\href {https://doi.org/10.1103/PhysRevLett.86.5188} {\bibfield
		{journal} {\bibinfo  {journal} {Phys. Rev. Lett.}\ }\textbf {\bibinfo
			{volume} {86}},\ \bibinfo {pages} {5188} (\bibinfo {year}
		{2001})}\BibitemShut {NoStop}%
	\bibitem [{\citenamefont {Li}\ and\ \citenamefont
		{Haldane}(2008)}]{li2008entanglement}%
	\BibitemOpen
	\bibfield  {author} {\bibinfo {author} {\bibfnamefont {H.}~\bibnamefont
			{Li}}\ and\ \bibinfo {author} {\bibfnamefont {F.~D.~M.}\ \bibnamefont
			{Haldane}},\ }\bibfield  {title} {\bibinfo {title} {Entanglement spectrum as
			a generalization of entanglement entropy: Identification of topological order
			in non-abelian fractional quantum hall effect states},\ }\href
	{https://doi.org/10.1103/PhysRevLett.101.010504} {\bibfield  {journal}
		{\bibinfo  {journal} {Phys. Rev. Lett.}\ }\textbf {\bibinfo {volume} {101}},\
		\bibinfo {pages} {010504} (\bibinfo {year} {2008})}\BibitemShut {NoStop}%
	\bibitem [{\citenamefont {Qi}\ \emph {et~al.}(2012)\citenamefont {Qi},
		\citenamefont {Katsura},\ and\ \citenamefont {Ludwig}}]{qi2012general}%
	\BibitemOpen
	\bibfield  {author} {\bibinfo {author} {\bibfnamefont {X.-L.}\ \bibnamefont
			{Qi}}, \bibinfo {author} {\bibfnamefont {H.}~\bibnamefont {Katsura}},\ and\
		\bibinfo {author} {\bibfnamefont {A.~W.~W.}\ \bibnamefont {Ludwig}},\
	}\bibfield  {title} {\bibinfo {title} {General relationship between the
			entanglement spectrum and the edge state spectrum of topological quantum
			states},\ }\href {https://doi.org/10.1103/PhysRevLett.108.196402} {\bibfield
		{journal} {\bibinfo  {journal} {Phys. Rev. Lett.}\ }\textbf {\bibinfo
			{volume} {108}},\ \bibinfo {pages} {196402} (\bibinfo {year}
		{2012})}\BibitemShut {NoStop}%
	\bibitem [{\citenamefont {Zou}\ and\ \citenamefont
		{Haah}(2016)}]{zou2016spurious}%
	\BibitemOpen
	\bibfield  {author} {\bibinfo {author} {\bibfnamefont {L.}~\bibnamefont
			{Zou}}\ and\ \bibinfo {author} {\bibfnamefont {J.}~\bibnamefont {Haah}},\
	}\bibfield  {title} {\bibinfo {title} {Spurious long-range entanglement and
			replica correlation length},\ }\href
	{https://doi.org/10.1103/PhysRevB.94.075151} {\bibfield  {journal} {\bibinfo
			{journal} {Phys. Rev. B}\ }\textbf {\bibinfo {volume} {94}},\ \bibinfo
		{pages} {075151} (\bibinfo {year} {2016})}\BibitemShut {NoStop}%
	\bibitem [{\citenamefont {Arildsen}\ \emph {et~al.}(2022)\citenamefont
		{Arildsen}, \citenamefont {Schuch},\ and\ \citenamefont
		{Ludwig}}]{arildsen2022entanglement}%
	\BibitemOpen
	\bibfield  {author} {\bibinfo {author} {\bibfnamefont {M.~J.}\ \bibnamefont
			{Arildsen}}, \bibinfo {author} {\bibfnamefont {N.}~\bibnamefont {Schuch}},\
		and\ \bibinfo {author} {\bibfnamefont {A.~W.}\ \bibnamefont {Ludwig}},\
	}\bibfield  {title} {\bibinfo {title} {Entanglement spectra of non-chiral
			topological (2+ 1)-dimensional phases with strong time-reversal breaking,
			li-haldane state counting, and peps},\ }\href@noop {} {\bibfield  {journal}
		{\bibinfo  {journal} {arXiv:2207.03246}\ } (\bibinfo {year}
		{2022})}\BibitemShut {NoStop}%
	\bibitem [{\citenamefont {Verstraete}\ \emph {et~al.}(2008)\citenamefont
		{Verstraete}, \citenamefont {Murg},\ and\ \citenamefont
		{Cirac}}]{verstraete2008matrix}%
	\BibitemOpen
	\bibfield  {author} {\bibinfo {author} {\bibfnamefont {F.}~\bibnamefont
			{Verstraete}}, \bibinfo {author} {\bibfnamefont {V.}~\bibnamefont {Murg}},\
		and\ \bibinfo {author} {\bibfnamefont {J.}~\bibnamefont {Cirac}},\ }\bibfield
	{title} {\bibinfo {title} {Matrix product states, projected entangled pair
			states, and variational renormalization group methods for quantum spin
			systems},\ }\href {https://doi.org/10.1080/14789940801912366} {\bibfield
		{journal} {\bibinfo  {journal} {Advances in Physics}\ }\textbf {\bibinfo
			{volume} {57}},\ \bibinfo {pages} {143} (\bibinfo {year} {2008})}\BibitemShut
	{NoStop}%
	\bibitem [{\citenamefont {Eisert}(2013)}]{eisert2013entanglement}%
	\BibitemOpen
	\bibfield  {author} {\bibinfo {author} {\bibfnamefont {J.}~\bibnamefont
			{Eisert}},\ }\bibfield  {title} {\bibinfo {title} {Entanglement and tensor
			network states},\ }\href@noop {} {\bibfield  {journal} {\bibinfo  {journal}
			{arXiv:1308.3318}\ } (\bibinfo {year} {2013})}\BibitemShut {NoStop}%
	\bibitem [{\citenamefont {Orús}(2014)}]{orus2014practical}%
	\BibitemOpen
	\bibfield  {author} {\bibinfo {author} {\bibfnamefont {R.}~\bibnamefont
			{Orús}},\ }\bibfield  {title} {\bibinfo {title} {A practical introduction to
			tensor networks: Matrix product states and projected entangled pair states},\
	}\href {https://doi.org/10.1016/j.aop.2014.06.013} {\bibfield  {journal}
		{\bibinfo  {journal} {Annals of Physics}\ }\textbf {\bibinfo {volume}
			{349}},\ \bibinfo {pages} {117–158} (\bibinfo {year} {2014})}\BibitemShut
	{NoStop}%
	\bibitem [{\citenamefont {Feldman}\ \emph {et~al.}(2022)\citenamefont
		{Feldman}, \citenamefont {Kshetrimayum}, \citenamefont {Eisert},\ and\
		\citenamefont {Goldstein}}]{feldman2022entanglement}%
	\BibitemOpen
	\bibfield  {author} {\bibinfo {author} {\bibfnamefont {N.}~\bibnamefont
			{Feldman}}, \bibinfo {author} {\bibfnamefont {A.}~\bibnamefont
			{Kshetrimayum}}, \bibinfo {author} {\bibfnamefont {J.}~\bibnamefont
			{Eisert}},\ and\ \bibinfo {author} {\bibfnamefont {M.}~\bibnamefont
			{Goldstein}},\ }\bibfield  {title} {\bibinfo {title} {Entanglement estimation
			in tensor network states via sampling},\ }\href
	{https://doi.org/10.1103/PRXQuantum.3.030312} {\bibfield  {journal} {\bibinfo
			{journal} {PRX Quantum}\ }\textbf {\bibinfo {volume} {3}},\ \bibinfo {pages}
		{030312} (\bibinfo {year} {2022})}\BibitemShut {NoStop}%
	\bibitem [{\citenamefont {Chen}\ \emph {et~al.}(2013)\citenamefont {Chen},
		\citenamefont {Gu}, \citenamefont {Liu},\ and\ \citenamefont
		{Wen}}]{chen2013symmetry}%
	\BibitemOpen
	\bibfield  {author} {\bibinfo {author} {\bibfnamefont {X.}~\bibnamefont
			{Chen}}, \bibinfo {author} {\bibfnamefont {Z.-C.}\ \bibnamefont {Gu}},
		\bibinfo {author} {\bibfnamefont {Z.-X.}\ \bibnamefont {Liu}},\ and\ \bibinfo
		{author} {\bibfnamefont {X.-G.}\ \bibnamefont {Wen}},\ }\bibfield  {title}
	{\bibinfo {title} {Symmetry protected topological orders and the group
			cohomology of their symmetry group},\ }\href
	{https://doi.org/10.1103/PhysRevB.87.155114} {\bibfield  {journal} {\bibinfo
			{journal} {Phys. Rev. B}\ }\textbf {\bibinfo {volume} {87}},\ \bibinfo
		{pages} {155114} (\bibinfo {year} {2013})}\BibitemShut {NoStop}%
	\bibitem [{\citenamefont {Calabrese}\ and\ \citenamefont
		{Cardy}(2004)}]{clabrese2004entanglement}%
	\BibitemOpen
	\bibfield  {author} {\bibinfo {author} {\bibfnamefont {P.}~\bibnamefont
			{Calabrese}}\ and\ \bibinfo {author} {\bibfnamefont {J.}~\bibnamefont
			{Cardy}},\ }\bibfield  {title} {\bibinfo {title} {Entanglement entropy and
			quantum field theory},\ }\href
	{https://doi.org/10.1088/1742-5468/2004/06/p06002} {\bibfield  {journal}
		{\bibinfo  {journal} {Journal of Statistical Mechanics: Theory and
				Experiment}\ }\textbf {\bibinfo {volume} {2004}},\ \bibinfo {pages} {P06002}
		(\bibinfo {year} {2004})}\BibitemShut {NoStop}%
	\bibitem [{\citenamefont {Chen}\ \emph {et~al.}(2011)\citenamefont {Chen},
		\citenamefont {Gu},\ and\ \citenamefont {Wen}}]{chen2011classification}%
	\BibitemOpen
	\bibfield  {author} {\bibinfo {author} {\bibfnamefont {X.}~\bibnamefont
			{Chen}}, \bibinfo {author} {\bibfnamefont {Z.-C.}\ \bibnamefont {Gu}},\ and\
		\bibinfo {author} {\bibfnamefont {X.-G.}\ \bibnamefont {Wen}},\ }\bibfield
	{title} {\bibinfo {title} {Classification of gapped symmetric phases in
			one-dimensional spin systems},\ }\href
	{https://doi.org/10.1103/PhysRevB.83.035107} {\bibfield  {journal} {\bibinfo
			{journal} {Phys. Rev. B}\ }\textbf {\bibinfo {volume} {83}},\ \bibinfo
		{pages} {035107} (\bibinfo {year} {2011})}\BibitemShut {NoStop}%
	\bibitem [{\citenamefont {Liu}\ and\ \citenamefont
		{Winter}(2022)}]{liu2020many}%
	\BibitemOpen
	\bibfield  {author} {\bibinfo {author} {\bibfnamefont {Z.-W.}\ \bibnamefont
			{Liu}}\ and\ \bibinfo {author} {\bibfnamefont {A.}~\bibnamefont {Winter}},\
	}\bibfield  {title} {\bibinfo {title} {Many-body quantum magic},\ }\href
	{https://doi.org/10.1103/PRXQuantum.3.020333} {\bibfield  {journal} {\bibinfo
			{journal} {PRX Quantum}\ }\textbf {\bibinfo {volume} {3}},\ \bibinfo {pages}
		{020333} (\bibinfo {year} {2022})}\BibitemShut {NoStop}%
	\bibitem [{\citenamefont {Fishman}\ \emph {et~al.}(2022)\citenamefont
		{Fishman}, \citenamefont {White},\ and\ \citenamefont
		{Stoudenmire}}]{fishman2020itensor}%
	\BibitemOpen
	\bibfield  {author} {\bibinfo {author} {\bibfnamefont {M.}~\bibnamefont
			{Fishman}}, \bibinfo {author} {\bibfnamefont {S.~R.}\ \bibnamefont {White}},\
		and\ \bibinfo {author} {\bibfnamefont {E.~M.}\ \bibnamefont {Stoudenmire}},\
	}\bibfield  {title} {\bibinfo {title} {{The ITensor Software Library for
				Tensor Network Calculations}},\ }\href@noop {} {\bibfield  {journal}
		{\bibinfo  {journal} {SciPost Phys. Codebases}\ }\textbf {\bibinfo {volume}
			{4}} (\bibinfo {year} {2022})}\BibitemShut {NoStop}%
	\bibitem [{\citenamefont {Teo}\ and\ \citenamefont
		{Kane}(2014)}]{teo2014luttinger}%
	\BibitemOpen
	\bibfield  {author} {\bibinfo {author} {\bibfnamefont {J.~C.~Y.}\
			\bibnamefont {Teo}}\ and\ \bibinfo {author} {\bibfnamefont {C.~L.}\
			\bibnamefont {Kane}},\ }\bibfield  {title} {\bibinfo {title} {From luttinger
			liquid to non-abelian quantum hall states},\ }\href
	{https://doi.org/10.1103/PhysRevB.89.085101} {\bibfield  {journal} {\bibinfo
			{journal} {Phys. Rev. B}\ }\textbf {\bibinfo {volume} {89}},\ \bibinfo
		{pages} {085101} (\bibinfo {year} {2014})}\BibitemShut {NoStop}%
	\bibitem [{\citenamefont {James}\ and\ \citenamefont
		{Konik}(2013)}]{james2013understanding}%
	\BibitemOpen
	\bibfield  {author} {\bibinfo {author} {\bibfnamefont {A.~J.~A.}\
			\bibnamefont {James}}\ and\ \bibinfo {author} {\bibfnamefont {R.~M.}\
			\bibnamefont {Konik}},\ }\bibfield  {title} {\bibinfo {title} {Understanding
			the entanglement entropy and spectra of 2d quantum systems through arrays of
			coupled 1d chains},\ }\href {https://doi.org/10.1103/PhysRevB.87.241103}
	{\bibfield  {journal} {\bibinfo  {journal} {Phys. Rev. B}\ }\textbf {\bibinfo
			{volume} {87}},\ \bibinfo {pages} {241103} (\bibinfo {year}
		{2013})}\BibitemShut {NoStop}%
	\bibitem [{\citenamefont {Cano}\ \emph {et~al.}(2015)\citenamefont {Cano},
		\citenamefont {Hughes},\ and\ \citenamefont
		{Mulligan}}]{cano2015interactions}%
	\BibitemOpen
	\bibfield  {author} {\bibinfo {author} {\bibfnamefont {J.}~\bibnamefont
			{Cano}}, \bibinfo {author} {\bibfnamefont {T.~L.}\ \bibnamefont {Hughes}},\
		and\ \bibinfo {author} {\bibfnamefont {M.}~\bibnamefont {Mulligan}},\
	}\bibfield  {title} {\bibinfo {title} {Interactions along an entanglement cut
			in $2+1\mathrm{D}$ abelian topological phases},\ }\href
	{https://doi.org/10.1103/PhysRevB.92.075104} {\bibfield  {journal} {\bibinfo
			{journal} {Phys. Rev. B}\ }\textbf {\bibinfo {volume} {92}},\ \bibinfo
		{pages} {075104} (\bibinfo {year} {2015})}\BibitemShut {NoStop}%
	\bibitem [{\citenamefont {Neupert}\ \emph {et~al.}(2014)\citenamefont
		{Neupert}, \citenamefont {Chamon}, \citenamefont {Mudry},\ and\ \citenamefont
		{Thomale}}]{neupert2014wire}%
	\BibitemOpen
	\bibfield  {author} {\bibinfo {author} {\bibfnamefont {T.}~\bibnamefont
			{Neupert}}, \bibinfo {author} {\bibfnamefont {C.}~\bibnamefont {Chamon}},
		\bibinfo {author} {\bibfnamefont {C.}~\bibnamefont {Mudry}},\ and\ \bibinfo
		{author} {\bibfnamefont {R.}~\bibnamefont {Thomale}},\ }\bibfield  {title}
	{\bibinfo {title} {Wire deconstructionism of two-dimensional topological
			phases},\ }\href {https://doi.org/10.1103/PhysRevB.90.205101} {\bibfield
		{journal} {\bibinfo  {journal} {Phys. Rev. B}\ }\textbf {\bibinfo {volume}
			{90}},\ \bibinfo {pages} {205101} (\bibinfo {year} {2014})}\BibitemShut
	{NoStop}%
	\bibitem [{\citenamefont {Lu}\ and\ \citenamefont
		{Vishwanath}(2012)}]{lu2012theory}%
	\BibitemOpen
	\bibfield  {author} {\bibinfo {author} {\bibfnamefont {Y.-M.}\ \bibnamefont
			{Lu}}\ and\ \bibinfo {author} {\bibfnamefont {A.}~\bibnamefont
			{Vishwanath}},\ }\bibfield  {title} {\bibinfo {title} {Theory and
			classification of interacting integer topological phases in two dimensions: A
			chern-simons approach},\ }\href {https://doi.org/10.1103/PhysRevB.86.125119}
	{\bibfield  {journal} {\bibinfo  {journal} {Phys. Rev. B}\ }\textbf {\bibinfo
			{volume} {86}},\ \bibinfo {pages} {125119} (\bibinfo {year}
		{2012})}\BibitemShut {NoStop}%
	\bibitem [{\citenamefont {Meng}\ \emph {et~al.}(2015)\citenamefont {Meng},
		\citenamefont {Neupert}, \citenamefont {Greiter},\ and\ \citenamefont
		{Thomale}}]{meng2015coupled}%
	\BibitemOpen
	\bibfield  {author} {\bibinfo {author} {\bibfnamefont {T.}~\bibnamefont
			{Meng}}, \bibinfo {author} {\bibfnamefont {T.}~\bibnamefont {Neupert}},
		\bibinfo {author} {\bibfnamefont {M.}~\bibnamefont {Greiter}},\ and\ \bibinfo
		{author} {\bibfnamefont {R.}~\bibnamefont {Thomale}},\ }\bibfield  {title}
	{\bibinfo {title} {Coupled-wire construction of chiral spin liquids},\ }\href
	{https://doi.org/10.1103/PhysRevB.91.241106} {\bibfield  {journal} {\bibinfo
			{journal} {Phys. Rev. B}\ }\textbf {\bibinfo {volume} {91}},\ \bibinfo
		{pages} {241106} (\bibinfo {year} {2015})}\BibitemShut {NoStop}%
	\bibitem [{\citenamefont {Gorohovsky}\ \emph {et~al.}(2015)\citenamefont
		{Gorohovsky}, \citenamefont {Pereira},\ and\ \citenamefont
		{Sela}}]{gorohovsky2015chiral}%
	\BibitemOpen
	\bibfield  {author} {\bibinfo {author} {\bibfnamefont {G.}~\bibnamefont
			{Gorohovsky}}, \bibinfo {author} {\bibfnamefont {R.~G.}\ \bibnamefont
			{Pereira}},\ and\ \bibinfo {author} {\bibfnamefont {E.}~\bibnamefont
			{Sela}},\ }\bibfield  {title} {\bibinfo {title} {Chiral spin liquids in
			arrays of spin chains},\ }\href {https://doi.org/10.1103/PhysRevB.91.245139}
	{\bibfield  {journal} {\bibinfo  {journal} {Phys. Rev. B}\ }\textbf {\bibinfo
			{volume} {91}},\ \bibinfo {pages} {245139} (\bibinfo {year}
		{2015})}\BibitemShut {NoStop}%
	\bibitem [{\citenamefont {Leviatan}\ and\ \citenamefont
		{Mross}(2020)}]{leviatan2020unification}%
	\BibitemOpen
	\bibfield  {author} {\bibinfo {author} {\bibfnamefont {E.}~\bibnamefont
			{Leviatan}}\ and\ \bibinfo {author} {\bibfnamefont {D.~F.}\ \bibnamefont
			{Mross}},\ }\bibfield  {title} {\bibinfo {title} {Unification of parton and
			coupled-wire approaches to quantum magnetism in two dimensions},\ }\href
	{https://doi.org/10.1103/PhysRevResearch.2.043437} {\bibfield  {journal}
		{\bibinfo  {journal} {Phys. Rev. Research}\ }\textbf {\bibinfo {volume}
			{2}},\ \bibinfo {pages} {043437} (\bibinfo {year} {2020})}\BibitemShut
	{NoStop}%
	\bibitem [{\citenamefont {Poilblanc}(2010)}]{poilblanc2010entanglement}%
	\BibitemOpen
	\bibfield  {author} {\bibinfo {author} {\bibfnamefont {D.}~\bibnamefont
			{Poilblanc}},\ }\bibfield  {title} {\bibinfo {title} {Entanglement spectra of
			quantum heisenberg ladders},\ }\href
	{https://doi.org/10.1103/PhysRevLett.105.077202} {\bibfield  {journal}
		{\bibinfo  {journal} {Phys. Rev. Lett.}\ }\textbf {\bibinfo {volume} {105}},\
		\bibinfo {pages} {077202} (\bibinfo {year} {2010})}\BibitemShut {NoStop}%
	\bibitem [{\citenamefont {Peschel}\ and\ \citenamefont
		{Chung}(2011)}]{peschel2011relation}%
	\BibitemOpen
	\bibfield  {author} {\bibinfo {author} {\bibfnamefont {I.}~\bibnamefont
			{Peschel}}\ and\ \bibinfo {author} {\bibfnamefont {M.-C.}\ \bibnamefont
			{Chung}},\ }\bibfield  {title} {\bibinfo {title} {On the relation between
			entanglement and subsystem hamiltonians},\ }\href
	{https://doi.org/10.1209/0295-5075/96/50006} {\bibfield  {journal} {\bibinfo
			{journal} {Europhysics Letters}\ }\textbf {\bibinfo {volume} {96}},\ \bibinfo
		{pages} {50006} (\bibinfo {year} {2011})}\BibitemShut {NoStop}%
	\bibitem [{\citenamefont {L\"auchli}\ and\ \citenamefont
		{Schliemann}(2012)}]{lauchli2012entanglement}%
	\BibitemOpen
	\bibfield  {author} {\bibinfo {author} {\bibfnamefont {A.~M.}\ \bibnamefont
			{L\"auchli}}\ and\ \bibinfo {author} {\bibfnamefont {J.}~\bibnamefont
			{Schliemann}},\ }\bibfield  {title} {\bibinfo {title} {Entanglement spectra
			of coupled $s=\frac{1}{2}$ spin chains in a ladder geometry},\ }\href
	{https://doi.org/10.1103/PhysRevB.85.054403} {\bibfield  {journal} {\bibinfo
			{journal} {Phys. Rev. B}\ }\textbf {\bibinfo {volume} {85}},\ \bibinfo
		{pages} {054403} (\bibinfo {year} {2012})}\BibitemShut {NoStop}%
	\bibitem [{\citenamefont {Mancini}\ \emph {et~al.}(2015)\citenamefont
		{Mancini}, \citenamefont {Pagano}, \citenamefont {Cappellini}, \citenamefont
		{Livi}, \citenamefont {Rider}, \citenamefont {Catani}, \citenamefont {Sias},
		\citenamefont {Zoller}, \citenamefont {Inguscio}, \citenamefont {Dalmonte},\
		and\ \citenamefont {Fallani}}]{mancini2015observation}%
	\BibitemOpen
	\bibfield  {author} {\bibinfo {author} {\bibfnamefont {M.}~\bibnamefont
			{Mancini}}, \bibinfo {author} {\bibfnamefont {G.}~\bibnamefont {Pagano}},
		\bibinfo {author} {\bibfnamefont {G.}~\bibnamefont {Cappellini}}, \bibinfo
		{author} {\bibfnamefont {L.}~\bibnamefont {Livi}}, \bibinfo {author}
		{\bibfnamefont {M.}~\bibnamefont {Rider}}, \bibinfo {author} {\bibfnamefont
			{J.}~\bibnamefont {Catani}}, \bibinfo {author} {\bibfnamefont
			{C.}~\bibnamefont {Sias}}, \bibinfo {author} {\bibfnamefont {P.}~\bibnamefont
			{Zoller}}, \bibinfo {author} {\bibfnamefont {M.}~\bibnamefont {Inguscio}},
		\bibinfo {author} {\bibfnamefont {M.}~\bibnamefont {Dalmonte}},\ and\
		\bibinfo {author} {\bibfnamefont {L.}~\bibnamefont {Fallani}},\ }\bibfield
	{title} {\bibinfo {title} {Observation of chiral edge states with neutral
			fermions in synthetic hall ribbons},\ }\href
	{https://doi.org/10.1126/science.aaa8736} {\bibfield  {journal} {\bibinfo
			{journal} {Science}\ }\textbf {\bibinfo {volume} {349}},\ \bibinfo {pages}
		{1510} (\bibinfo {year} {2015})}\BibitemShut {NoStop}%
	\bibitem [{\citenamefont {Petrescu}\ and\ \citenamefont
		{Le~Hur}(2015)}]{petrescu2015chiral}%
	\BibitemOpen
	\bibfield  {author} {\bibinfo {author} {\bibfnamefont {A.}~\bibnamefont
			{Petrescu}}\ and\ \bibinfo {author} {\bibfnamefont {K.}~\bibnamefont
			{Le~Hur}},\ }\bibfield  {title} {\bibinfo {title} {Chiral mott insulators,
			meissner effect, and laughlin states in quantum ladders},\ }\href
	{https://doi.org/10.1103/PhysRevB.91.054520} {\bibfield  {journal} {\bibinfo
			{journal} {Phys. Rev. B}\ }\textbf {\bibinfo {volume} {91}},\ \bibinfo
		{pages} {054520} (\bibinfo {year} {2015})}\BibitemShut {NoStop}%
	\bibitem [{\citenamefont {Cornfeld}\ and\ \citenamefont
		{Sela}(2015)}]{cornfeld2015chiral}%
	\BibitemOpen
	\bibfield  {author} {\bibinfo {author} {\bibfnamefont {E.}~\bibnamefont
			{Cornfeld}}\ and\ \bibinfo {author} {\bibfnamefont {E.}~\bibnamefont
			{Sela}},\ }\bibfield  {title} {\bibinfo {title} {Chiral currents in
			one-dimensional fractional quantum hall states},\ }\href
	{https://doi.org/10.1103/PhysRevB.92.115446} {\bibfield  {journal} {\bibinfo
			{journal} {Phys. Rev. B}\ }\textbf {\bibinfo {volume} {92}},\ \bibinfo
		{pages} {115446} (\bibinfo {year} {2015})}\BibitemShut {NoStop}%
	\bibitem [{\citenamefont {Calvanese~Strinati}\ \emph
		{et~al.}(2017)\citenamefont {Calvanese~Strinati}, \citenamefont {Cornfeld},
		\citenamefont {Rossini}, \citenamefont {Barbarino}, \citenamefont {Dalmonte},
		\citenamefont {Fazio}, \citenamefont {Sela},\ and\ \citenamefont
		{Mazza}}]{strinati2017laughlin}%
	\BibitemOpen
	\bibfield  {author} {\bibinfo {author} {\bibfnamefont {M.}~\bibnamefont
			{Calvanese~Strinati}}, \bibinfo {author} {\bibfnamefont {E.}~\bibnamefont
			{Cornfeld}}, \bibinfo {author} {\bibfnamefont {D.}~\bibnamefont {Rossini}},
		\bibinfo {author} {\bibfnamefont {S.}~\bibnamefont {Barbarino}}, \bibinfo
		{author} {\bibfnamefont {M.}~\bibnamefont {Dalmonte}}, \bibinfo {author}
		{\bibfnamefont {R.}~\bibnamefont {Fazio}}, \bibinfo {author} {\bibfnamefont
			{E.}~\bibnamefont {Sela}},\ and\ \bibinfo {author} {\bibfnamefont
			{L.}~\bibnamefont {Mazza}},\ }\bibfield  {title} {\bibinfo {title}
		{Laughlin-like states in bosonic and fermionic atomic synthetic ladders},\
	}\href {https://doi.org/10.1103/PhysRevX.7.021033} {\bibfield  {journal}
		{\bibinfo  {journal} {Phys. Rev. X}\ }\textbf {\bibinfo {volume} {7}},\
		\bibinfo {pages} {021033} (\bibinfo {year} {2017})}\BibitemShut {NoStop}%
	\bibitem [{\citenamefont {Calvanese~Strinati}\ \emph
		{et~al.}(2019)\citenamefont {Calvanese~Strinati}, \citenamefont {Sahoo},
		\citenamefont {Shtengel},\ and\ \citenamefont
		{Sela}}]{strinati2019pretopological}%
	\BibitemOpen
	\bibfield  {author} {\bibinfo {author} {\bibfnamefont {M.}~\bibnamefont
			{Calvanese~Strinati}}, \bibinfo {author} {\bibfnamefont {S.}~\bibnamefont
			{Sahoo}}, \bibinfo {author} {\bibfnamefont {K.}~\bibnamefont {Shtengel}},\
		and\ \bibinfo {author} {\bibfnamefont {E.}~\bibnamefont {Sela}},\ }\bibfield
	{title} {\bibinfo {title} {Pretopological fractional excitations in the
			two-leg flux ladder},\ }\href {https://doi.org/10.1103/PhysRevB.99.245101}
	{\bibfield  {journal} {\bibinfo  {journal} {Phys. Rev. B}\ }\textbf {\bibinfo
			{volume} {99}},\ \bibinfo {pages} {245101} (\bibinfo {year}
		{2019})}\BibitemShut {NoStop}%
	\bibitem [{\citenamefont {Bridgeman}\ and\ \citenamefont
		{Chubb}(2017)}]{bridgeman2017hand}%
	\BibitemOpen
	\bibfield  {author} {\bibinfo {author} {\bibfnamefont {J.~C.}\ \bibnamefont
			{Bridgeman}}\ and\ \bibinfo {author} {\bibfnamefont {C.~T.}\ \bibnamefont
			{Chubb}},\ }\bibfield  {title} {\bibinfo {title} {Hand-waving and
			interpretive dance: an introductory course on tensor networks},\ }\href
	{https://doi.org/10.1088/1751-8121/aa6dc3} {\bibfield  {journal} {\bibinfo
			{journal} {Journal of Physics A: Mathematical and Theoretical}\ }\textbf
		{\bibinfo {volume} {50}},\ \bibinfo {pages} {223001} (\bibinfo {year}
		{2017})}\BibitemShut {NoStop}%
	\bibitem [{\citenamefont {Hastings}(2014)}]{hastings2014notes}%
	\BibitemOpen
	\bibfield  {author} {\bibinfo {author} {\bibfnamefont {M.}~\bibnamefont
			{Hastings}},\ }\bibfield  {title} {\bibinfo {title} {Notes on some questions
			in mathematical physics and quantum information},\ }\href@noop {} {\bibfield
		{journal} {\bibinfo  {journal} {arXiv:1404.4327}\ } (\bibinfo {year}
		{2014})}\BibitemShut {NoStop}%
\end{thebibliography}
\end{document}